\documentclass[a4paper,10pt,twocolumn,showpacs,notitlepage,nofootinbib,floatfix,superscriptaddress,prd]{revtex4-1}
\usepackage{graphicx}
\usepackage{amsfonts}
\usepackage{amsmath}
\usepackage{float}
\usepackage{amssymb}

\begin{document}

\title{Influence of temperature dependent shear viscosity on
  elliptic flow at back- and forward rapidities in
  ultrarelativistic heavy-ion collisions}

\begin{abstract}
  We explore the influence of a temperature-dependent shear viscosity
  over entropy density ratio $\eta_s/s$ on the azimuthal anisotropies
  $v_2$ and $v_4$ of hadrons at various rapidities. We find that in
  Au~+~Au collisions at full Relativistic Heavy Ion Collider energy,
  $\protect\sqrt{s_{NN}}=200$ GeV, the flow anisotropies are dominated
  by hadronic viscosity at all rapidities, whereas in Pb~+~Pb
  collisions at the Large Hadron Collider energy,
  $\protect\sqrt{s_{NN}}=2760$ GeV, the flow coefficients are affected
  by the viscosity in both the plasma and hadronic phases at
  midrapidity, but the further away from midrapidity, the more
  dominant the hadronic viscosity becomes. We find that the centrality and
  rapidity dependence of the elliptic and quadrangular flows can help
  to distinguish different parametrizations of $(\eta_s/s)(T)$. We
  also find that at midrapidity the flow harmonics are almost
  independent of the decoupling criterion, but show some sensitivity
  to the criterion at back- and forward rapidities.
\end{abstract}

\date{\today}

\author{E.\ Moln\'ar}
\affiliation{MTA-DE Particle Physics Research Group, H-4010 Debrecen,
             P.O.\ Box 105, Hungary} 
\affiliation{Frankfurt Institute for Advanced Studies, Ruth-Moufang-Strasse 1,
             D-60438 Frankfurt am Main, Germany}

\author{H.\ Holopainen}
\affiliation{Frankfurt Institute for Advanced Studies, Ruth-Moufang-Strasse 1,
             D-60438 Frankfurt am Main, Germany}

\author{P.\ Huovinen}
\affiliation{Frankfurt Institute for Advanced Studies, Ruth-Moufang-Strasse 1,
             D-60438 Frankfurt am Main, Germany} 
\affiliation{Institut f\"ur Theoretische Physik,
             Johann Wolfgang Goethe-Universit\"at,
             Max-von-Laue-Strasse 1, D-60438 Frankfurt am Main, Germany}

\author{H.\ Niemi}
\affiliation{Department of Physics, University of Jyv\"askyl\"a,
             P.O.\ Box 35 (YFL), FI-40014 University of Jyv\"askyl\"a, Finland}
\affiliation{Helsinki Institute of Physics, P.O.\ Box 64, FI-00014 University
of Helsinki, Finland}

\maketitle

\section{Introduction}

Determining the transport properties of the quark-gluon plasma (QGP)
formed in ultrarelativistic nuclear collisions~\cite{experiments} is
nowadays one of the main goals in high-energy nuclear physics.
Fluid-dynamical models indicate a very low shear viscosity to
entropy density ratio $\eta_s/s$\footnote{In this work $\eta_s$
  denotes the coefficient of shear viscosity, $\eta_{\textrm{ch}}$ the
  pseudorapidity, and $\eta$ the space-time rapidity.}, when tuned to
reproduce the azimuthal anisotropies of the transverse momentum
distributions of observed hadrons. For recent reviews see, for example,
Refs.~\cite{Heinz:2013th,Gale:2013da,Huovinen:2013wma}.  The values
favored by state-of-the-art calculations are in the vicinity of the
conjectured lower limit for shear viscosity, $\eta_s/s = 1/(4\pi)$,
based on the anti--de Sitter/conformal field theory (AdS/CFT)
correspondence~\cite{Policastro:2001yc}. For example, the values found
in Ref.~\cite{Gale:2012rq} are $\eta_s/s = 0.12$ for collisions at the
Relativistic Heavy-Ion Collider (RHIC) at Brookhaven National
Laboratory, and $\eta_s/s = 0.2$ at the Large Hadron Collider (LHC) at
CERN.

The values quoted above were obtained using a constant $\eta_s/s$
ratio during the entire evolution of the system. For a physical system
$\eta_s/s$ depends at least on temperature~\cite{Csernai:2006zz} and
on baryon density~\cite{Denicol:2013nua}. A constant value of
$\eta_s/s$ represents only an effective average over the entire
space-time evolution of the system. The slightly larger effective
$\eta_s/s$ obtained for collisions at the LHC, \emph{i.e.}, at larger
collision energy, thus may be interpreted as an indication of the
temperature dependence of $\eta_s/s$~\cite{Song:2011qa,Heinz:2013wva}.
Unfortunately, extracting the temperature dependence of $\eta_s/s$ from
the experimental data is a challenging problem.

In our previous works \cite{Niemi:2011ix,SQM,Niemi:2012ry}, we have
studied the consequences of relaxing the assumption of a constant
$\eta_s/s$. We found that the relevant temperature region where the
shear viscosity affects the elliptic flow most varies with the
collision energy. At RHIC the most relevant region is around and below
the QCD transition temperature, while for higher collision energies
the temperature region above the transition becomes more and more
important. To constrain the temperature dependence of $\eta_s/s$
better, it would thus be necessary to find observables which are
sensitive to the shear viscosity at different stages of the evolution
of a single collision.

In this work we relax the assumption of boost invariance of our
earlier works, solve the evolution equations numerically in all three
dimensions, and study whether the azimuthal anisotropies have similar
dependence on $(\eta_s/s)(T)$ at all rapidities. If not, the
measurements of $v_n$ at back- and forward rapidities could bring
further constraints to $(\eta_s/s)(T)$.

We also approach the problem of extracting the temperature dependence
of $\eta_s/s$ in a fashion similar to Ref.~\cite{Song:2011qa}: We tune
different parametrizations to reproduce the anisotropies at one
collision energy and centrality, and check whether anisotropies at
different centralities, rapidities, and collision energies can
distinguish between these parametrizations.

Furthermore, we check the sensitivity of our results to different
decoupling criteria. To this end we carry out the calculations using a
dynamical freeze-out criterion, \emph{i.e.}, freeze-out at constant
Knudsen
number~\cite{Niemi:2014wta,Holopainen:2012id,Holopainen:2013jna}, and
compare the results to those obtained using the conventional
freeze-out at constant temperature.

In the following we describe the structure and freeze-out in our
(3+1)-dimensional dissipative fluid dynamical model in
Sec.~\ref{fluid}, and the parameters in our calculations in
Sec.~\ref{parameters}.  Section~\ref{results} contains the comparison
of our results with experimental data, while in
Secs.~\ref{rapidity_v2} and~\ref{dynamical_FO} we discuss whether it
is possible to distinguish the details of different parametrizations
of $(\eta_s/s)(T)$, as well as the effects of a dynamical freeze-out
criterion. We summarize our results in Sec.~\ref{conclusions}.

Specific details of the fluid-dynamical equations are relegated to
Appendix~\ref{3+1d_equations}. The numerical algorithm and details of
our implementation, and the numerical accuracy of our code, are
discussed in Appendices~\ref{shasta}, and~\ref{shasta_accuracy},
respectively.

In this work we use natural units $\hbar=c=k=1$.

\section{Fluid dynamics}
  \label{fluid}

\subsection{Equations of motion}
  \label{fluid_dynamics}

Relativistic fluid dynamics corresponds to the local conservation of
energy-momentum and net-charge currents (if any),
\begin{equation}
\partial _{\mu }T^{\mu \nu }=0,\quad \partial _{\mu }N_i^{\mu }=0,
 \label{cons_eqs}
\end{equation}
where $T^{\mu \nu }$ is the energy-momentum tensor
and $N_i^{\mu }$ are the net-charge four-currents.

These macroscopic fields can be decomposed with respect to the fluid
flow velocity defined by Landau and Lifshitz \cite{Landau_book},
$u^{\mu }=T^{\mu \nu }u_{\nu }/e$, as
\begin{eqnarray}
T^{\mu\nu} & = & eu^{\mu}u^{\nu} - P\Delta^{\mu\nu} 
                + \pi^{\mu\nu},  \label{T_munu_general} \\
N^{\mu}_i   & = & n_i u^{\mu} + V_i^{\mu} ,
\label{N_mu}
\end{eqnarray}
where $e = T^{\mu\nu}u_{\mu}u_{\nu}$, and $n_i = N_i^{\mu}u_{\mu }$
are the energy and net-charge densities in the local rest frame,
respectively, $P = -T^{\mu\nu}\Delta_{\mu\nu}/3$ is the isotropic
pressure, and $V_i^{\mu} = N_i^{\alpha}\Delta_{\alpha}^{\mu }$
are the charge diffusion currents. The shear-stress tensor,
$\pi^{\mu\nu} = T^{\left\langle \mu \nu \right\rangle }$, is the
traceless and orthogonal part of the energy-momentum tensor.
With the $(+,-,-,-)$ convention for the metric tensor $g^{\mu\nu}$, 
the projection tensor is $\Delta^{\mu\nu} = g^{\mu\nu} - u^{\mu}u^{\nu}$.
The angular brackets $\left\langle {  }\right\rangle $ denote an
operator leading to the symmetric, traceless, and orthogonal to the
flow velocity part of a tensor: $T^{\left\langle \mu \nu \right\rangle } = 
\left[ \frac{1}{2}\left( \Delta_{\alpha}^{\mu}\Delta_{\beta}^{\nu} 
                        +\Delta_{\beta}^{\mu}\Delta_{\alpha}^{\nu}\right)
     - \frac{1}{3}\Delta^{\mu\nu}\Delta_{\alpha\beta}\right] T^{\alpha\beta}$.

Landau's matching condition allows one to associate the rest-frame
densities with their equilibrium values, $e = e_0(T,\{\mu_i\})$, and
$n_i = n_{i,0}(T,\{\mu_j\})$. The difference between the isotropic and
equilibrium pressures defines the so-called bulk viscosity, $\Pi = P-P_0$.

Equations~(\ref{T_munu_general}) and (\ref{N_mu}) can be closed by
providing an equation of state (EoS), together with the equations
determining the evolution of dissipative quantities $\pi^{\mu\nu}$,
$\Pi$, and $V_i^\mu$. These quantities represent the dissipative
forces in the system as well as deviations from the local thermal
equilibrium.  In the Navier-Stokes approximation they are linearly
proportional to the gradients of velocity and temperature, with
proportionality coefficients for shear viscosity
$\eta_s(T,\{\mu_i\})$, bulk viscosity $\zeta(T,\{\mu_i\})$, and charge
diffusion $\kappa_i(T,\{\mu_j\})$ quantifying the transport properties
of the matter.

It is well known that the bulk viscosity coefficient of a relativistic
gas is about three orders of magnitude smaller than its shear
viscosity coefficient, and vanishes in the ultrarelativistic
limit~\cite{Cercignani_book}. However, it is still important for
relativistic systems around phase transitions; therefore, even if the
bulk viscosity is negligible in the QGP-phase, it may be large near
and below the phase transition~\cite{Torrieri:2007fb}.  A large bulk
viscosity at those stages may or may not have a significant effect on
the observables~\cite{Monnai:2009ad,Song:2009rh,Bozek:2011ua,Dusling:2011fd,Noronha-Hostler:2013gga,Noronha-Hostler:2014dqa}. Since
disentangling the effects of shear and bulk on the observed spectra is
difficult, and beyond the scope of this work, we adopt the approach of
Ref.~\cite{Song:2009rh}. We assume that bulk viscosity is large only
in the vicinity of the QCD phase transition but due to the critical
slowing down its effect is so small that it can be safely ignored.

At midrapidity the matter formed in ultrarelativistic collisions at
RHIC and at the LHC is to a good approximation net-baryon free, and
thus in boost-invariant calculations it has been an excellent
approximation to neglect all conserved charges. Since in this study we
want to investigate the back- and forward rapidity regions of the
system where net-baryon density is finite, in principle we should
include the net-baryon current and baryon charge diffusion in the
description of the system. However, the baryon charge diffusion in a
QGP as well as in a hadron gas is largely unknown at the moment.
Also, at low values of net-baryon density where the lattice QCD
results~\cite{Borsanyi:2011sw,Bazavov:2012jq} can be used, the effect
of the finite density on the EoS is small~\cite{Huovinen:2012xm}.
Therefore, to simplify the description of the system, and to allow us
to concentrate solely on the effects of shear viscosity on the
spectra, we ignore the finite baryon charge in the fluid as well. Thus
we are left with the shear-stress tensor $\pi^{\mu\nu}$ as the only
dissipative quantity in the system.

In so-called second-order or causal fluid-dynamical theories by
M\"{u}ller and by Israel and
Stewart~\cite{Muller_67,Muller_99,Israel:1979wp} the dissipative
quantities fulfill certain coupled relaxation equations. Here we
recall the relaxation equation for the shear-stress tensor obtained
from the relativistic Boltzmann
equation~\cite{Denicol:2012cn,Denicol:2012es,Molnar:2013lta},
\begin{align}
 \tau _{\pi}D\pi^{\mu\nu} 
  & = 2\eta_{s}\sigma^{\mu\nu} - \pi^{\mu\nu}
    - \tau_{\pi}\left( \pi^{\lambda\mu}u^{\nu} +\pi^{\lambda\nu}u^{\mu}\right) 
               Du_{\lambda }  \notag \\
  & - \delta_{\pi\pi}\pi^{\mu\nu}\theta 
    - \tau_{\pi\pi}\pi_{\lambda}^{\left\langle \mu \right.}
                 \sigma^{\left.\nu \right\rangle \lambda} \notag \\
  & + 2\tau_{\pi}\pi_{\lambda}^{\left\langle \mu \right.}
                \omega^{\left.\nu \right\rangle \lambda}
    + \varphi_{7}\pi_{\lambda}^{\left\langle \mu\right.}\pi^{\left. \nu \right\rangle\lambda}\ .
  \label{relax_shear}
\end{align}
Here $\tau_{\pi}$ is the shear-stress relaxation time,
$D\pi^{\mu\nu} = u^\alpha \pi^{\mu\nu}_{;\alpha}$ denotes the time
derivative, $\theta$ the expansion rate, $\sigma^{\mu \nu}$ the
shear tensor and $\omega^{\mu \nu}$ the vorticity. The other
coefficients can be calculated self-consistently from microscopic
theory and, for example, in case of an ultrarelativistic massless
Boltzmann gas we obtain, in the 14-moment approximation,
$\tau_{\pi}=\frac{5}{3}\lambda_{mfp}$, 
$\delta_{\pi\pi} = \left( 4/3\right) \tau_{\pi}$,
$\ \tau_{\pi\pi} = \left( 10/7\right) \tau_{\pi}$, while
$\varphi_{7} = \left( 9/70\right)/P_{0}$, where $\lambda_{mfp}$ is the
mean free path between collisions. For QCD these coefficients are
mostly unknown; however, for high-temperature QCD matter the
coefficients given above may be acceptable as a first approximation.

For the sake of simplicity we ignore the last two terms in
Eq.~(\ref{relax_shear}). This is justified since the relative
contribution of the $\varphi_7$ coefficient was shown to be negligible
compared to the others~\cite{Molnar:2013lta}. Similarly, we have
observed that the term proportional to the vorticity has little effect
on the overall evolution of the system, and is thus omitted from the
final calculations shown here.


\subsection{The freeze-out stage}
  \label{freeze_out}

During the fluid-dynamical evolution the system cools and dilutes due
to the expansion, and consequently the microscopic rescattering rate
of particles, $\Gamma \sim n\sigma \simeq \lambda_{mfp}^{-1}$,
decreases, until the rescatterings cease and particles stream freely
toward detectors. The transition from an (almost) equilibrated fluid
to free-streaming particles is a gradual process, but since
implementing such a gradual process into a fluid-dynamical description
is very complicated~\cite{Csernai:2005ht,Akkelin:2008eh}, it is
usually assumed to take place on an infinitesimally thin space-time
layer, on the so-called freeze-out surface. Therefore the total number
of particles crossing the surface $\Sigma$, with a normal vector
$d^{3}\Sigma_{\mu}$ pointing outward, leads to the following invariant
distribution of particles emitted from the fluid, known as the
Cooper-Frye formula~\cite{Cooper:1974mv}:
\begin{equation}
 E\frac{d^3N}{d^3p} = \int_{\Sigma}d^3\Sigma_{\mu}(x)\,p^{\mu}f(x,p)\,,
   \label{Cooper-Frye}
\end{equation}
where $p^{\mu } = \left( E,\mathbf{p}\right)$ denotes the
four-momentum, while $f\left( x,p\right)$ is the phase-space
distribution function of particles on the surface.

To apply the Cooper-Frye formula, we need an appropriate criterion for
choosing the surface $\Sigma$. Since scattering rates strongly depend
on temperature, the usual approach is to assume the freeze-out to take
place on a surface of constant temperature or energy density. However,
it has been argued that it would be more physical to assume that the
freeze-out happens when the average scattering rate is roughly equal
to the expansion rate of the system~\cite{Bondorf:1978kz}.

This latter, so-called dynamical freeze-out, criterion can be
expressed in terms of the Knudsen number, $\text{Kn}$, which is the
ratio of a characteristic microscopic time or length scale, such as
$\lambda_{mfp}$, and a characteristic macroscopic scale of the fluid,
such as the inverse of the local gradients,
$L^{-1}\approx \partial_{\mu}$. In terms of the Knudsen number the
dynamical freeze-out criterion is $\text{Kn}\approx 1$, which has
occasionally been used in ideal fluid
calculations~\cite{Hung:1997du,Eskola:2007zc,Holopainen:2012id,Holopainen:2013jna},
but for viscous fluids it is more appropriate to use the relaxation
time(s) of dissipative quantity(ies) as the microscopic scale, since
they appear naturally in the evolution equations for dissipative
quantities~\cite{Denicol:2012cn}.

In most of our calculations we use the conventional
constant-temperature freeze-out, but to evaluate how sensitive our
results are to the particular freeze-out criterion, and to the
freeze-out description in general, we also do the calculations
assuming freeze-out at constant Knudsen number. We take the relaxation
time of shear stress, $\tau_\pi$, as the microscopic scale, and the
inverse of the expansion rate of the system, $\theta^{-1}$, as the
macroscopic scale. Thus we get a local Knudsen number of
\begin{equation}
 \text{Kn} = \tau_{\pi}\theta \ .
\end{equation}
Since the Knudsen number can be evaluated in many different
ways~\cite{Niemi:2014wta}, we do not insist on freeze-out at
$\text{Kn} = 1$, but treat the freeze-out Knudsen number as a free
parameter chosen to reproduce rapidity and $p_T$ distributions of
experimental data. To avoid pathologies encountered in
Refs.~\cite{Niemi:2014wta,Holopainen:2012id}, we also require that the
dynamical freeze-out takes place below a temperature of $T=180$ MeV
and above $T = 80$ MeV.

To evaluate the distributions on the freeze-out surface, we assume
that the distribution of particles for each species $i$, \emph{i.e.},
$f_{i}(x,p)$, is given by the well-known Grad's 14-moment ansatz,
which includes corrections $\delta f_{i}$ (shear viscosity only) to
the local equilibrium distribution function as
\begin{equation}
f_i(x,p)\equiv f_{0i} + \delta f_{i} 
  = f_{0i}\left[ 1+ \left(1 \mp \tilde{f}_{0i}\right)
                   \frac{p_{i}^{\mu}p_{i}^{\nu}\pi_{\mu\nu}}
                       {2 T^2\left( e+p\right) }\right] ,
\label{f_final}
\end{equation}
where $f_{0i}$ is the local equilibrium distribution function,
\begin{equation}
 f_{0i}\left( x,p\right) = \frac{g_i}{\left( 2\pi \right) ^{3}}
                          \left[ \exp\left( \frac{p_{i}^{\mu }u_{\mu }-\mu_i }
                                                 {T}\right) \pm 1\right]^{-1},
\label{eq_distribution}
\end{equation}
and $\tilde{f}_{0i} = (2\pi)^3 f_{0i}/g_i$.
We also include the contribution from all strong and electromagnetic
two- and three-particle decays of the hadronic resonances up to a mass
of $2$ GeV mass to the final particle distributions.

The flow anisotropies are defined from a Fourier decomposition of the
particle spectra as
\begin{equation}
E\frac{d^3 N}{d^3 p} = \frac{d^2 N}{2 \pi p_T d p_T dy_{p} } 
\left(1 + 2 \sum_{n=1}^\infty v_n\cos n(\phi- \Psi_n) \right),
\end{equation}
where $y_{p}=\frac{1}{2}\ln \left[ (p^{0}+p^{z})/(p^{0}-p^{z})\right]$
is the rapidity of the particle, $p_T=\sqrt{p^2_x + p^2_y}$ its
transverse momenta, and $\Psi_n$ is the event plane for coefficient
$v_n$.  The Fourier coefficients $v_n=v_n(p_T,y_{p})$ are the
differential flow components.  In this work the differential and
integrated $v_n$ are calculated using the event-plane method.


\section{Parameters}
\label{parameters}

We mostly implement the parametrization used in
Refs.~\cite{Niemi:2011ix,Niemi:2012ry}, but retune the parameter
values, and generalize it for a (3+1)-dimensional non-boost-invariant
case.

\subsection{Equation of state}

For the EoS we use the $s95p$-PCE-v1 parametrization of lattice QCD
results at zero net-baryon density~\cite{Huovinen:2009yb}. The
high-temperature part of the EoS is given by the hotQCD
Collaboration~\cite{Cheng:2007jq,Bazavov:2009zn} and it is smoothly
connected to the low-temperature part described as a hadron resonance
gas, where resonances up to a mass of $2$ GeV are included. The hadronic
part includes a chemical freeze-out at $T_{\mathrm{chem}}=150$ MeV
where all stable particle ratios are fixed~\cite{Bebie:1991ij,
  Hirano:2002ds, Huovinen:2007xh}. Since the construction of the EoS
assumes that the entropy per particle is conserved after chemical
freeze-out, the small (approximately $1\%$) entropy increase during
the viscous hydrodynamical evolution below $T_\mathrm{chem}$ leads to
a small increase in particle yields too.

\subsection{Transport coefficients}

\begin{figure} 
\vspace{-0.1cm}
\includegraphics[width=8cm]{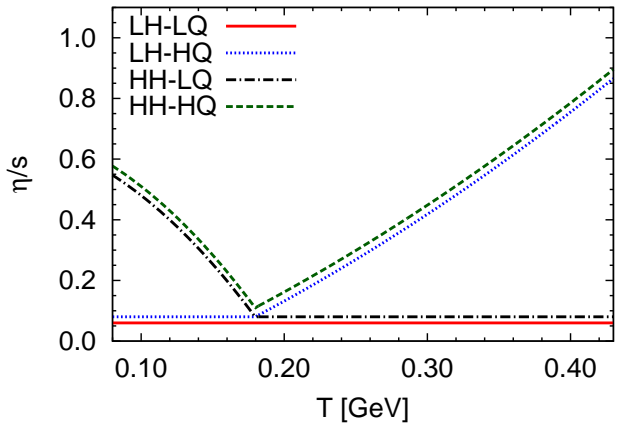} 
\vspace{-0.3cm}
\caption{{\protect\small (Color online) Different parametrizations of
    $\eta_s/s$ as a function of temperature. The LH-LQ line has been
    shifted downward and the HH-HQ upward for better visibility.}}
\label{fig:etapers}
\end{figure}

As in our earlier works~\cite{Niemi:2011ix,SQM,Niemi:2012ry}, we use
four different parametrizations of the temperature-dependent shear
viscosity over entropy ratio, see Fig.~\ref{fig:etapers}:
\begin{itemize}
\item LH-LQ, in which $(\eta_s/s)(T)=0.08$ for all temperatures;

\item LH-HQ, in which $(\eta_s/s)(T)=0.08$ for the hadronic
  phase, while above $T_{tr}$ the viscosity to entropy ratio increases
  according to
\begin{eqnarray}
\lefteqn{(\eta_s/s)(T)_{\mathrm{QGP}}}\\ 
    & = & -0.289+0.288\,\frac{T}{T_{\mathrm{tr}}}
          +0.0818\left( \frac{T}{T_{\mathrm{tr}}}\right)^2 \, ; \nonumber
\end{eqnarray}

\item HH-LQ, in which, in the hadronic phase below $T_{tr}$, 
\begin{eqnarray}
\lefteqn{(\eta_s/s)(T)_{\mathrm{HRG}}}\\
    & = &  0.681-0.0594\,\frac{T}{T_{\mathrm{tr}}}
          -0.544\left( \frac{T}{T_{\mathrm{tr}}}\right)^2 \, , \nonumber
\end{eqnarray}
while in the QGP-phase $(\eta_s/s)(T)=0.08$;

\item HH-HQ, in which we use $(\eta_s/s)(T)_{\mathrm{HRG}}$ and
  $(\eta_s/s)(T)_{\mathrm{QGP}}$ for the hadronic and QGP
  phases, respectively.
\end{itemize}

Unless stated otherwise, the value of $\eta_s/s$ at the transition
temperature, $T_{tr}=180$ MeV, is $(\eta_s/s)(T_{tr})=0.08$. This is a
close approximation to the lower bound conjectured in the framework of
the AdS/CFT correspondence~\cite{Policastro:2001yc}. For all
parametrizations the relaxation time for the shear-stress tensor is
\begin{equation}
\tau _{\pi} = 5\frac{\eta_s }{e+p}. \label{eq:relax}
\end{equation}
For the sake of comparison, we also do the calculations using zero
shear viscosity, \emph{i.e.}, an ideal fluid.

\subsection{The initial state}
  \label{initial_conditions}

In this work we ignore the effects of event-by-event
fluctuations~\cite{Luzum:2013yya,Pang:2012he}, and generalize a simple
optical Glauber model~\cite{Miller:2007ri} for a non-boost-invariant
initial state. In different variants of the Glauber model the initial
energy density in the transverse plane at midrapidity and at initial
time $\tau_0$ is given as a function of the density of binary
collisions, $n_{BC}(x,y,b)$, wounded nucleons, $n_{WN}(x,y,b)$, or
both:
\begin{equation}
 e_T\left( \tau _{0},x,y,b\right) 
  = C_{e}(\tau_0)\ f\left( n_{BC},n_{WN}\right) ,
  \label{glauber}
\end{equation}
where the normalization constant $C_{e}(\tau_0)$ is selected to
reproduce the multiplicity measured in central collisions, and $b$ is
the impact parameter of the collision. In the following we use our
BCfit parametrization~\cite{Niemi:2011ix,Niemi:2012ry}, where the
energy density depends solely on the number of binary collisions:
\begin{equation}
  f_{BC}\left( n_{BC},n_{WN}\right) = n_{BC} + c_{1}n_{BC}^{2} + c_{2}n_{BC}^{3},
\label{BC_fit}
\end{equation}
and the coefficients $c_1$ and $c_2$ are chosen to reproduce the
observed centrality dependence of multiplicity.

In the optical Glauber model, the density of binary collisions on the
transverse plane is calculated from
\begin{equation}
 n_{BC}\left( x,y,b\right) 
   = \sigma_{NN}T_A\left( x+b/2,y\right) T_B\left( x-b/2,y\right) , \label{n_BC}
\end{equation}
where $\sigma_{NN}$ is the total nucleon-nucleon inelastic cross
section, and $T_{A/B}$ is the nuclear thickness function. As a cross
section we use $\sigma_{NN}=42$ mb at
RHIC~\cite{Miller:2007ri,Alver:2008aq}, and $\sigma_{NN}=64$ mb at the
LHC~\cite{Abelev:2013qoq}. As usual, we define the thickness function
as
\begin{equation}
 T_A\left( x,y\right) = \int_{-\infty }^{\infty }dz\,\rho_A\left( x,y,z\right) ,
\end{equation}
where $\rho_A$ is the Woods-Saxon nuclear density distribution,
\begin{equation}
\rho_A\left( \mathbf{r}\right) 
  = \frac{\rho_0}{1+ \exp \left[ \left(r-R_{A}\right) /d\right] },
\end{equation}
and $\rho_{0}= 0.17$ fm$^{-3}$ is the ground-state nuclear density,
and $d=0.54$ fm is the surface thickness. The nuclear radii $R_A$ are
calculated from $R_{A}=1.12A^{1/3}-0.86/A^{1/3}$, which gives
$R_{Au}\simeq 6.37$ fm and $R_{Pb}\simeq 6.49$ fm ($A_{Au}=197$ and
$A_{Pb}=208$).

Unfortunately there are very few theoretical constraints for the
longitudinal structure of the initial state, since even the most
sophisticated approaches to calculate the initial state from basic
principles~\cite{ipglasma,Paatelainen:2013eea} are restricted to
midrapidity. Here we follow the simple approaches shown in
Refs.~\cite{Hirano:2001eu,Nonaka:2006yn,Schenke:2010nt}, and in a
similar fashion assume longitudinal scaling flow, $v_z = z/t$,
\emph{i.e.}, $v_{\eta} = 0$, and a constant energy density
distribution around midrapidity~\cite{Bjorken:1982qr}, followed by
exponential tails in both back- and forward directions. We parametrize
the longitudinal energy density distribution as
\begin{equation}
 e_L\left( \eta \right) 
  = \exp \left( -2c_{\eta }
                \sqrt{1+\frac{\left( |\eta|-\eta_0\right)^2}{2c_\eta\sigma_\eta^2}
                \Theta\left( |\eta |-\eta_0\right) }
              + 2c_\eta\right) ,  \label{f_eta}
\end{equation}
where $\eta =\frac{1}{2}\ln \left[ (t+z)/(t-z)\right] $ is the
space-time rapidity, and $\Theta(x)$ the Heaviside step function. Thus
the normalized energy density distribution is
\begin{equation}
  e\left( \tau_{0},x,y,\eta ,b\right) 
   = e_{T}\left( \tau_{0},x,y,b\right) e_L\left( \eta \right).
  \label{Glauber_energy_density}
\end{equation}
We are aware that there are more sophisticated approaches in the
literature~\cite{Adil:2005qn,Bozek:2011ua,Hirano:2012kj,Vovchenko:2013viu},
but since attempts to create more plausible longitudinal structures
easily lead to a rapidity distribution of $v_2$ which strongly
deviates from the observed one~\cite{Hirano:2001eu}, we leave the
detailed study of the longitudinal structures for a later work.

Due to entropy production in dissipative fluids, the different
parametrizations of $\eta_s /s$ lead to different entropy production
and therefore different final multiplicity of hadrons. Because most of
the entropy is produced during the early stages of the expansion when
the longitudinal gradients are largest \cite{Dumitru:2007qr}, it is
sufficient to adjust initial densities according to the entropy
produced in the partonic phase. Further entropy production during the
hadronic evolution turns out to represent only a small contribution in
the final multiplicities and it is not corrected in our calculations.

At RHIC, we used as maximum energy density,
$e_0 = e(\tau_0,0,0,0)$, for
\begin{itemize}
\item an ideal fluid: $e_{0}=17.0$ GeV/fm$^{3}$,

\item LH-LQ and HH-LQ: $e_{0}=15.8$ GeV/fm$^{3}$,

\item LH-HQ and HH-HQ: $e_{0}=14.9$ GeV/fm$^{3}$,
\end{itemize}
while at the LHC we used for
\begin{itemize}
\item an ideal fluid: $e_{0}=57.5$ GeV/fm$^{3}$,

\item LH-LQ and HH-LQ: $e_{0}=54.5$ GeV/fm$^{3}$,

\item LH-HQ and HH-HQ: $e_{0}=49.5$ GeV/fm$^{3}$.
\end{itemize}
Note that these values are smaller than the ones given in
Refs.~\cite{Niemi:2011ix,Niemi:2012ry}. The main reason is that we
used different data to fit the centrality dependence, and chose to fit
the multiplicity as a function of centrality class, not as a function
of number of participants, as was done in
Refs.~\cite{Niemi:2011ix,Niemi:2012ry}. This leads to different values
of $c_1$ and $c_2$ parameters, and, consequently, the maximum density
in a head-on collision (which practically never happens) is different
even if the energy density at midrapidity at impact parameters $b > 2$
fm is almost identical.

The parameters controlling the centrality dependence, $c_1$ and $c_2$
in Eq.~(\ref{BC_fit}), are $c_1 = -0.035$ fm$^{-2}$, and $c_2=0.00034$
fm$^{-4}$ at RHIC, and $c_1 = -0.02$ fm$^{-2}$ and $c_2 = 0.000175$
fm$^{-4}$ at the LHC. The parameters in Eq.~(\ref{f_eta}) defining the
longitudinal structure are $c_{\eta} = 4$ at RHIC and $c_{\eta} = 2$
at the LHC, while $\eta_{0} = 2.0$ for the constant rapidity plateau
for both. The width of the rapidity distribution is $\sigma_{\eta} =
1.0$ at RHIC and $\sigma _{\eta }=1.8$ at the LHC. The average impact
parameters in each centrality class are given in Table \ref{tab:ini}.

\begin{table}
\begin{center}
\begin{tabular}{|c|c|c|}
\hline
Centrality (\%) & RHIC $b$ (fm) & LHC $b$ (fm) \\ \hline
0-5 & $2.24$ & $2.32$ \\ \hline
5-10 & $4.09$ & $4.24$ \\ \hline
10-20 & $5.78$ & $5.99$ \\ \hline
20-30 & $7.49$ & $7.76$ \\ \hline
30-40 & $8.87$ & $9.19$ \\ \hline
40-50 & $10.06$ & $10.43$ \\ \hline
\end{tabular}
\end{center}
\par
\vspace*{-0.1cm} \vspace*{-0.3cm}
\caption{{\protect\small The average impact parameter $b$ in each centrality
class at RHIC and the LHC.}}
\label{tab:ini}
\end{table}

If not stated otherwise the fluid-dynamical evolution is started at
$\tau_0 = 1$ fm/$c$ proper time. The initial values for the transverse
fluid velocity and shear-stress tensor are always set to zero. The
value of the decoupling temperature or Knudsen number is indicated in
the figures.

To obtain the final particle distributions we use the framework
described in Ref.~\cite{Holopainen:2010gz}. Thus we sample particle
distributions to create ``events'' even if we are not doing
event-by-event calculations, but use conventional averaged initial
states. The particle spectra and other measurables at RHIC are
obtained as an average over $N_{ev}=100\,000$ events, where the sampling
is done over $p_T = (0, 5.4) $ GeV and $\eta_{ch}=(-6.6,6.6)$ with
$N_{p_T} = 36$ and $N_{\eta_{ch}}=22$ bins. At the LHC the particle
multiplicity is $\simeq 2.5$ times larger than at RHIC; hence we
average over $N_{ev}=40\,000$ events.


\section{Results and comparisons to data}
  \label{results}

\subsection{Au+Au at $\protect\sqrt{s_{NN}}=200$ GeV at RHIC} 
  \label{RHIC_results}

\begin{figure} 
\vspace{-0.1cm} 
\includegraphics[width=8.5cm]{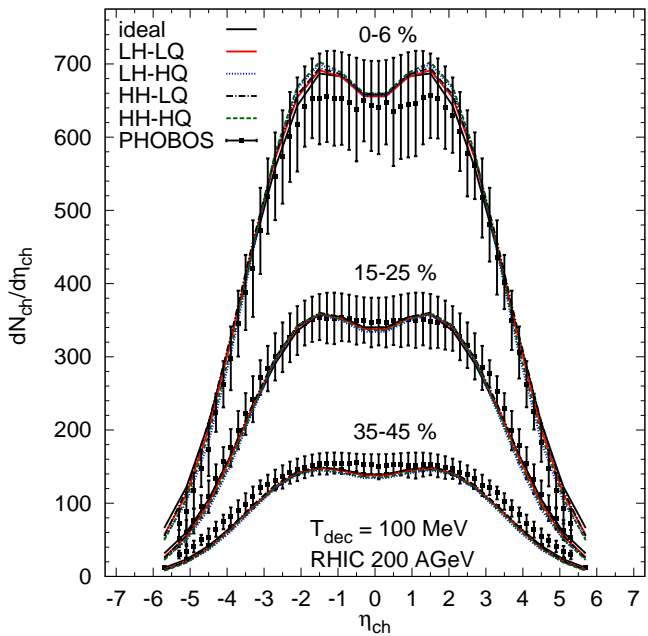}
\vspace{-0.3cm}
\caption{{\protect\small (Color online) The charged particle
    pseudorapidity distribution $dN_{ch}/d\protect\eta_{ch}$.
    Experimental data are from the PHOBOS Collaboration
    \protect\cite{Back:2002wb}. }}
\label{fig:charged_hadron_y_RHIC}
\end{figure}

We fix the parameters characterizing the initial state,
Eqs.~(\ref{glauber}), (\ref{BC_fit}), and~(\ref{f_eta}), by comparison
to the PHOBOS charged particle pseudorapidity distribution,
$dN_{ch}/d\eta_{ch}$, at various centralities~\cite{Back:2002wb}. We
present our results in Fig.~\ref{fig:charged_hadron_y_RHIC}, where the
calculations are shown for 0--5\% centrality, and for the average of
10--20\% and 20--30\% as well as 30--40\% and 40--50\%
centralities. This is in order to facilitate a comparison to the data
taken at 0--6\%, 15--25\% and, 35--45\% centralities. As required, the
final multiplicity and pseudorapidity distribution are well reproduced
at all centralities for all parametrizations of the
temperature-dependent shear viscosity to entropy density ratio.  Here
we once again stress the importance of fixing the initial energy
density to compensate for the entropy production for different
$\eta_s/s$ parametrizations.  Otherwise, for fixed initial densities,
the larger the effective viscosity, the larger the entropy production
and thus the final multiplicity.

The kinetic freeze-out temperature, $T_{dec}$, affects the charged
particle pseudorapidity distribution very weakly. We have chosen
$T_{dec} = 100$ MeV by comparison to the pion, kaon and proton
$p_T$ spectra measured by the PHENIX Collaboration~\cite{Adler:2003cb},
and checked that if we use $T_{dec} = 140$ MeV, the pseudorapidity
distributions are still within error bars, and the change is on the
same level as the differences due to different viscosities shown in
Fig.~\ref{fig:charged_hadron_y_RHIC}. Such a weak dependence is not
surprising: It is well known that in a chemically frozen system pion
$p_T$ distributions are weakly sensitive to the kinetic freeze-out
temperature~\cite{Hirano:2005wx}. We now observe similar behavior in
the longitudinal direction.

\begin{figure}
\vspace{-0.1cm} 
\includegraphics[width=8.5cm]{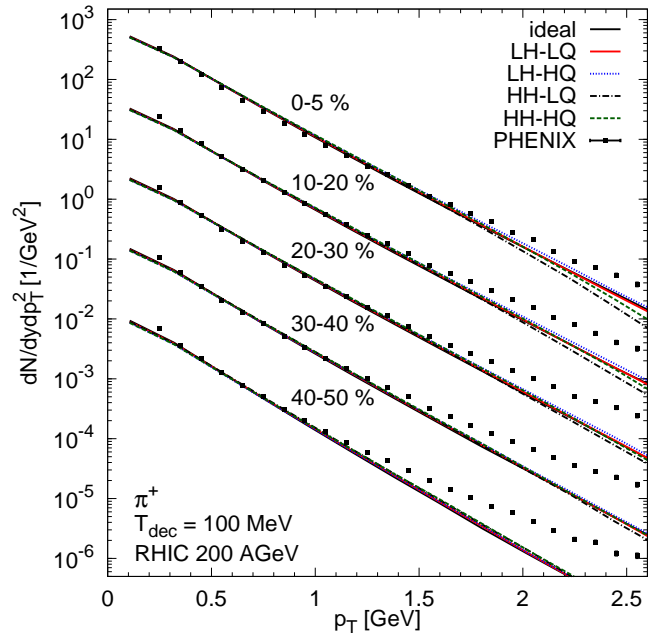} 
\vspace{-0.3cm}
\caption{{\protect\small (Color online) Transverse momentum spectra of
    positive pions at RHIC. Experimental data are from the PHENIX
    Collaboration \protect\cite{Adler:2003cb}. }}
\label{fig:pion_spectra_RHIC}
\end{figure}
\begin{figure}
\vspace{-0.1cm} 
\includegraphics[width=8.5cm]{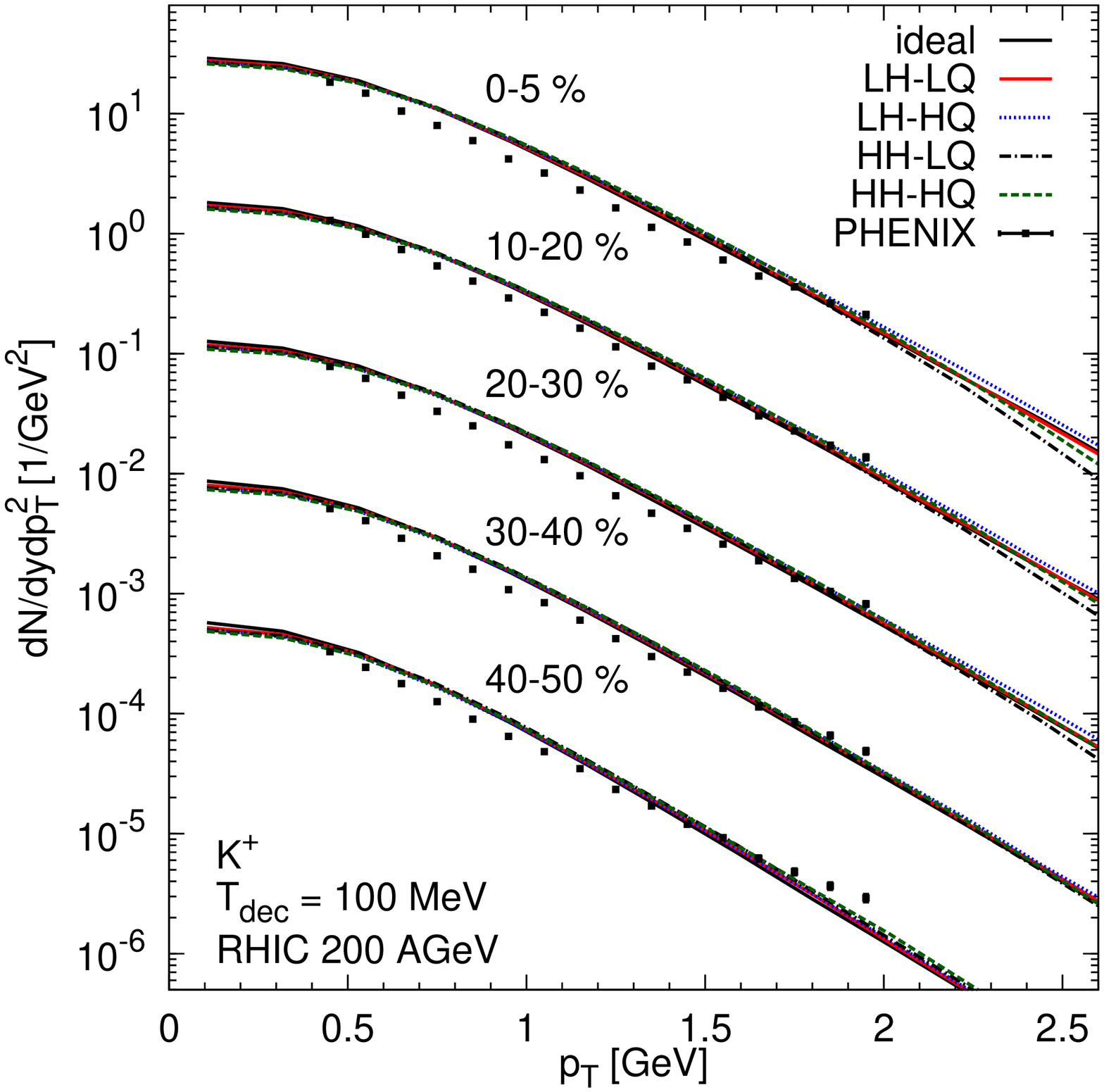} 
\vspace{-0.3cm}
\caption{{\protect\small (Color online) Transverse momentum spectra of
    positive kaons at RHIC. Experimental data are from the PHENIX
    Collaboration \protect\cite{Adler:2003cb}. }}
\label{fig:kaon_spectra_RHIC}
\end{figure}
\begin{figure}
\vspace{-0.1cm} 
\includegraphics[width=8.5cm]{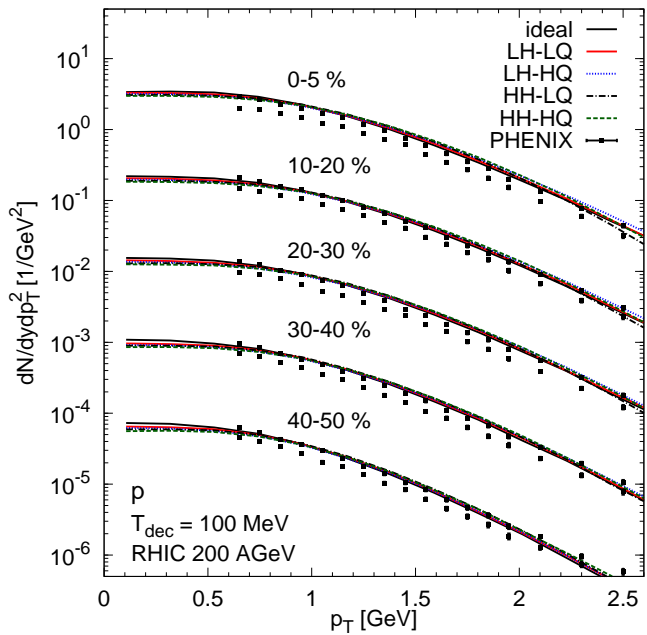} 
\vspace{-0.3cm}
\caption{{\protect\small (Color online) Transverse momentum spectra of
    protons at RHIC. Experimental data for protons (upper) and
    antiprotons (lower) are from the PHENIX Collaboration
    \protect\cite{Adler:2003cb}. }}
\label{fig:proton_spectra_RHIC}
\end{figure}

In Figs.~\ref{fig:pion_spectra_RHIC}, \ref{fig:kaon_spectra_RHIC},
and~\ref{fig:proton_spectra_RHIC} we present the $p_T$ spectra of
positive pions, kaons, and protons, respectively, corresponding to
centrality classes, 0--5\%, 10--20\%$(\times 10^{-1})$,
20--30\%$(\times 10^{-2})$, 30--40\%$(\times 10^{-3})$, and
40--50\%$(\times 10^{-4})$, Here the multiplicative factors are
applied (to both theoretical and experimental points) for better
visibility. The experimental data are from the PHENIX
Collaboration~\cite{Adler:2003cb}.

As seen before in viscous calculations (\emph{e.g.},~in
Ref.~\cite{Niemi:2012ry}), the slopes of pion spectra are reasonably
well reproduced up to $p_T \simeq 1.5$ GeV for semicentral collisions,
but the agreement recedes with increasing impact parameter. The kaon
yields are overpredicted at all centralities, whereas the fit to
proton spectra is slightly better than the fit to kaons. Since we do
not include a finite baryochemical potential in our calculation, we
are consistently overestimating the yields of heavy particles, which
might imply the need for even lower chemical freeze-out temperature.

The pion spectra become flatter with increasing freeze-out
temperature; hence for example for $T_{dec}=140$ MeV the theoretical
calculations are in a better agreement at larger momenta, but 
overestimate the spectra around $p_T \sim 1$ GeV. The slope of the
proton spectra become steeper with increasing freeze-out temperature
as well, and thus $T_{dec} = 100$ MeV provides the best compromise.

As expected, after the initial densities are fixed to reproduce the
yield, the slopes are practically unaffected by the different
$\eta_s/s$ parametrizations, and the corresponding $\delta f_i$ in
each case represents only a small correction compared to the thermal
spectra.

\begin{figure} 
\vspace{-0.1cm} 
\includegraphics[width=8.5cm]{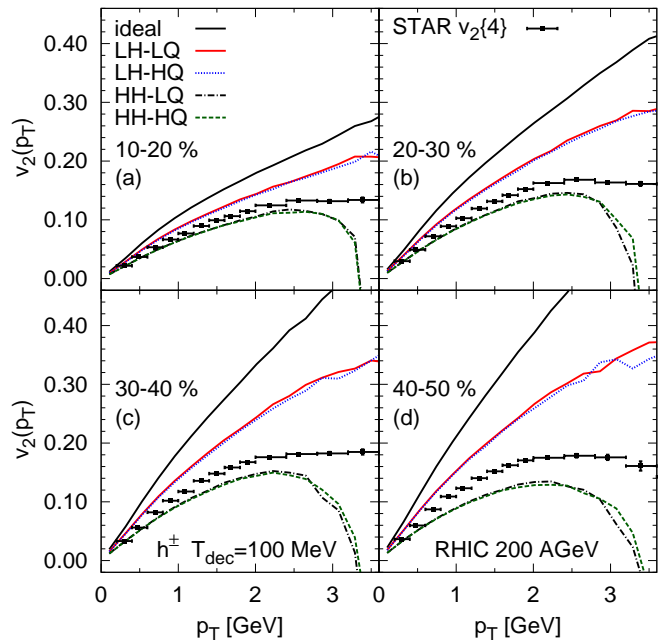} 
\vspace{-0.3cm}
\caption{{\protect\small (Color online) Charged hadron $v_2(p_T)$ at
    RHIC. Experimental data are from the STAR Collaboration
    \protect\cite{Bai}. }}
\label{fig:charged_hadron_v2_RHIC}
\end{figure}
\begin{figure} 
\vspace{-0.1cm} 
\includegraphics[width=8.5cm]{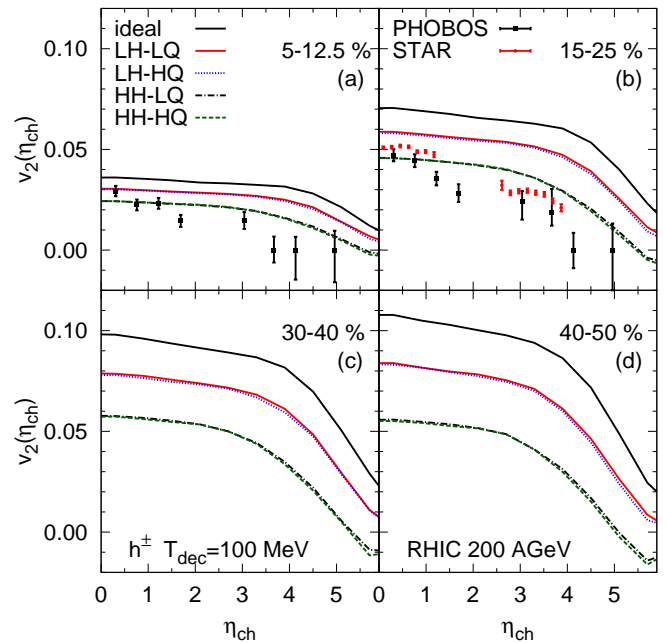}
\vspace{-0.3cm}
\caption{{\protect\small (Color online) Charged hadron
    $v_2(\protect\eta_{ch})$ at RHIC. Experimental data are
    from the PHOBOS \protect\cite{Back:2004mh} and STAR
    \protect\cite{Adams:2004bi} Collaborations. }}
\label{fig:charged_hadron_v2_y_RHIC}
\end{figure}

In Figs.~\ref{fig:charged_hadron_v2_RHIC}
and~\ref{fig:charged_hadron_v2_y_RHIC} the elliptic flow coefficient
$v_2$ at various centralities is shown as a function of transverse
momentum $p_T$, and pseudorapidity $\eta_{ch}$. In
Fig.~\ref{fig:charged_hadron_v2_RHIC} the experimental data are from
the STAR Collaboration~\cite{Bai}, whereas in
Fig.~\ref{fig:charged_hadron_v2_y_RHIC} the average of 0--5\% and
10--20\% and of 10--20\% and 20--30\% events are compared to the data
from the PHOBOS Collaboration for 3--15\% and 15--25\% centrality
classes~\cite{Back:2004mh}, and to the STAR Collaboration data in the
15--25\% centrality class~\cite{Adams:2004bi}.

As expected, the $p_T$ differential elliptic flow coefficient shows
the behavior reported in Refs.~\cite{Niemi:2011ix,Niemi:2012ry}: At
RHIC the elliptic flow coefficient is very sensitive to viscosity in
the hadronic phase but independent of the high-temperature
parametrization of the viscosity. The same observation also holds for
the rapidity-dependent elliptic flow coefficient at all centrality
classes. The dissipative reduction of $v_2$ is quite independent of
rapidity, and thus we cannot reproduce the shape of $v_2(\eta_{ch})$
very well. On the other hand, slightly larger hadronic viscosity would
further reduce $v_2$, and our result would be very close to the ideal
fluid + UrQMD hybrid calculation of Ref.~\cite{Hirano:2005xf}.

\begin{figure} 
\vspace{-0.1cm} 
\includegraphics[width=8.5cm]{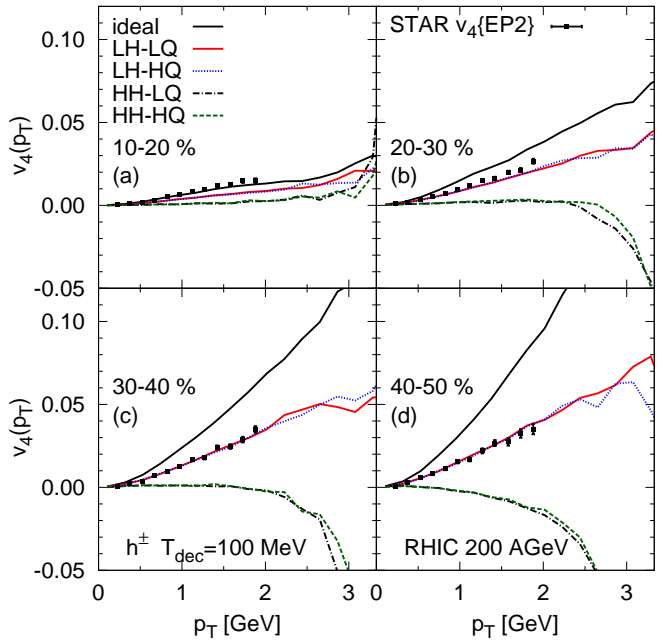} 
\vspace{-0.3cm}
\caption{{\protect\small (Color online) Charged hadron $v_4(p_T)$ at
    RHIC. Experimental data are from the STAR Collaboration
    \cite{Adams:2004bi}.}}
\label{fig:charged_hadron_v4_RHIC}
\end{figure}
\begin{figure} 
\vspace{-0.1cm} 
\includegraphics[width=8.5cm]{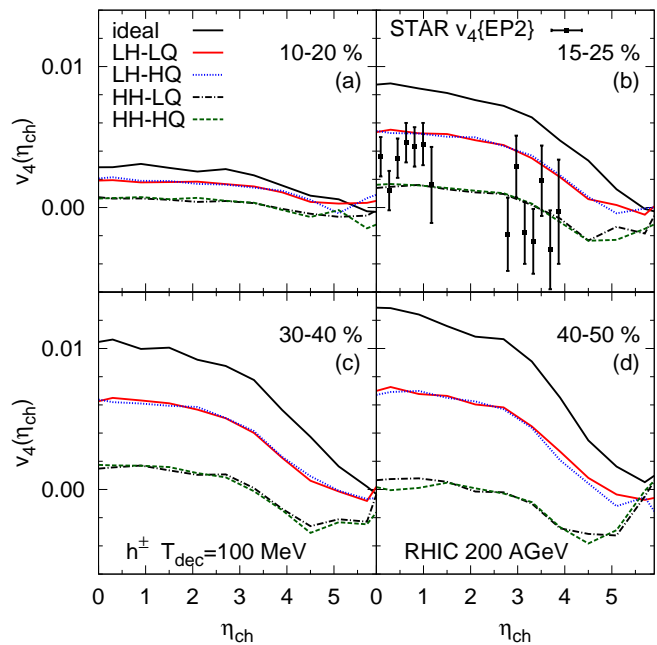}
\vspace{-0.3cm} 
\caption{{\protect\small (Color online) Charged hadron
    $v_4(\protect\eta_{ch})$ at RHIC. Experimental data are
    from the STAR Collaboration \cite{Adams:2004bi}. }}
\label{fig:charged_hadron_v4_y_RHIC}
\end{figure}

Similarly the $v_4(p_T)$ and $v_4(\eta_{ch})$ of charged hadrons in
different centrality classes are compared to the experimental data
from the STAR Collaboration~\cite{Adams:2004bi} in
Figs.~\ref{fig:charged_hadron_v4_RHIC}
and~\ref{fig:charged_hadron_v4_y_RHIC}. The $v_4$ coefficient, both as
a function of transverse momentum and as a function of pseudorapidity,
complies with the previously made observations about the elliptic flow
coefficient. As we have reported earlier~\cite{SQM,Niemi:2012ry},
$v_4$ is sensitive to viscosity at even later stages of the evolution
than $v_2$, and a large hadronic viscosity is sufficient to turn
$v_4(p_T)$ negative at quite low $p_T$. The comparison of
Figs.~\ref{fig:charged_hadron_v2_y_RHIC}
and~\ref{fig:charged_hadron_v4_y_RHIC} also shows the well-known fact
that the larger the value of $n$, the stronger the viscous suppression
of $v_n$~\cite{Alver:2010dn,Schenke:2011bn}. Viscosity has only a weak
effect on the shapes of $v_2(\eta_{ch})$ and $v_4(\eta_{ch})$, but
quite interestingly the effect on the shapes is different for
different coefficients: Increasing viscosity makes the (approximate)
plateau in $v_2(\eta_{ch})$ narrower but in $v_4(\eta_{ch})$ wider.

From Fig.~\ref{fig:charged_hadron_v4_RHIC} it is apparent that the
$v_4(p_T)$ data favor the parametrizations with low hadronic viscosity
unlike $v_2(p_T)$. However, we have to remember that the experimental
data were obtained using different methods for $v_2$ and $v_4$,
\emph{i.e.}, four-particle cumulant and mixed harmonic event-plane
methods, whereas we use the event-plane method to evaluate all the
harmonics. Another uncertainty is that event-by-event fluctuations
cause a sizable fraction of $v_4$, but they are not included in our
study. Thus we advise against drawing any conclusions about the
favored $(\eta_s/s)(T)$ from this particular result.

\subsection{Pb+Pb at $\protect\sqrt{s_{NN}}=2760$ GeV at the LHC}
  \label{LHC_results}

As at RHIC, we use the pseudorapidity distribution of charged
particles to fix the initialization, and the $p_T$ distributions of
identified particles to fix the kinetic freeze-out temperature.

\begin{figure} 
\vspace{-0.1cm} 
\includegraphics[width=8.5cm]{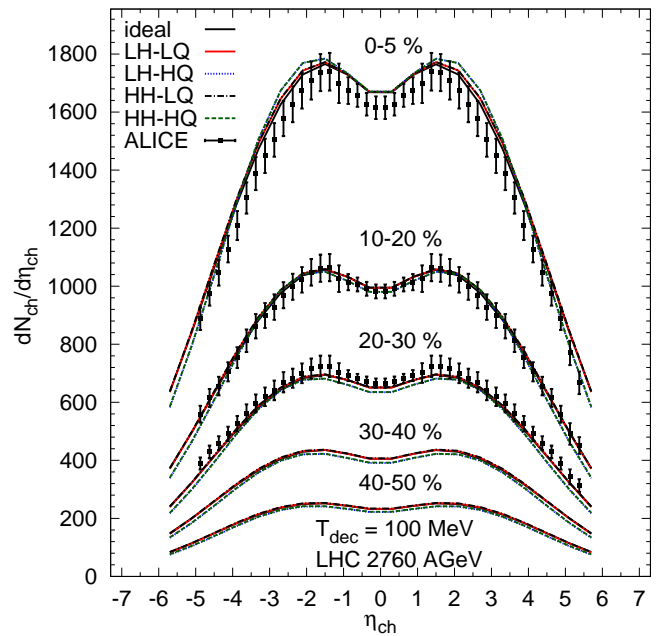} 
\vspace{-0.3cm}
\caption{{\protect\small (Color online) The charged particle
    pseudorapidity distribution $dN_{ch}/d\protect\eta_{ch}$ at the
    LHC. Experimental data are from the ALICE Collaboration
    \protect\cite{Abbas:2013bpa}. }}
\label{fig:charged_hadron_y_LHC}
\end{figure}

In Fig.~\ref{fig:charged_hadron_y_LHC} the charged particle
pseudorapidity distribution $dN_{ch}/d\eta_{ch}$ for different
centrality bins are compared to the experimental data from the ALICE
Collaboration~\cite{Abbas:2013bpa}.  The pseudorapidity distribution
of charged particles reasonably matches the data for all centrality
classes given in the figure. Similarly as for RHIC we slightly
overshoot the experimental results at the LHC for the most central
collisions while we undershoot the peripheral ones. Moreover, as
observed before, the pseudorapidity distributions of charged particles
are insensitive to the chosen freeze-out temperature.

\begin{figure}
\vspace{-0.1cm} 
\includegraphics[width=8.5cm]{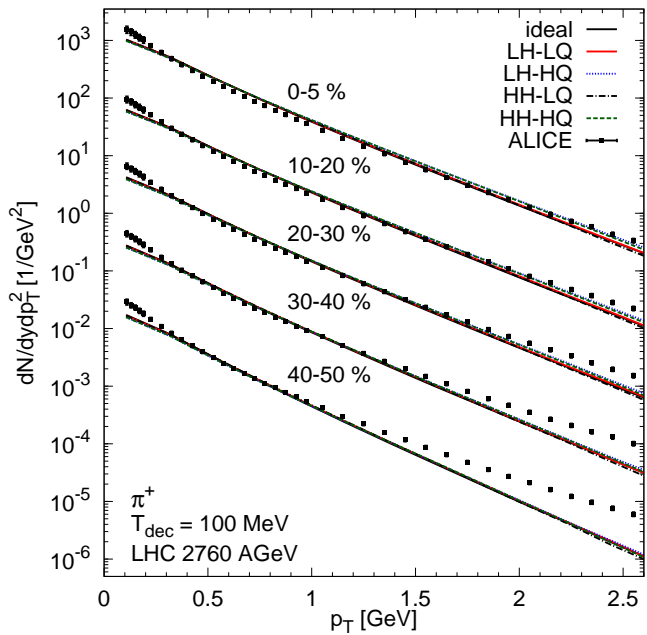} 
\vspace{-0.3cm}
\caption{{\protect\small (Color online) Transverse momentum spectra of
    positive pions at the LHC. Experimental data are from the ALICE
    Collaboration \protect\cite{Abelev:2013vea}. }}
\label{fig:pion_spectra_LHC}
\end{figure}
\begin{figure}
\vspace{-0.1cm} 
\includegraphics[width=8.5cm]{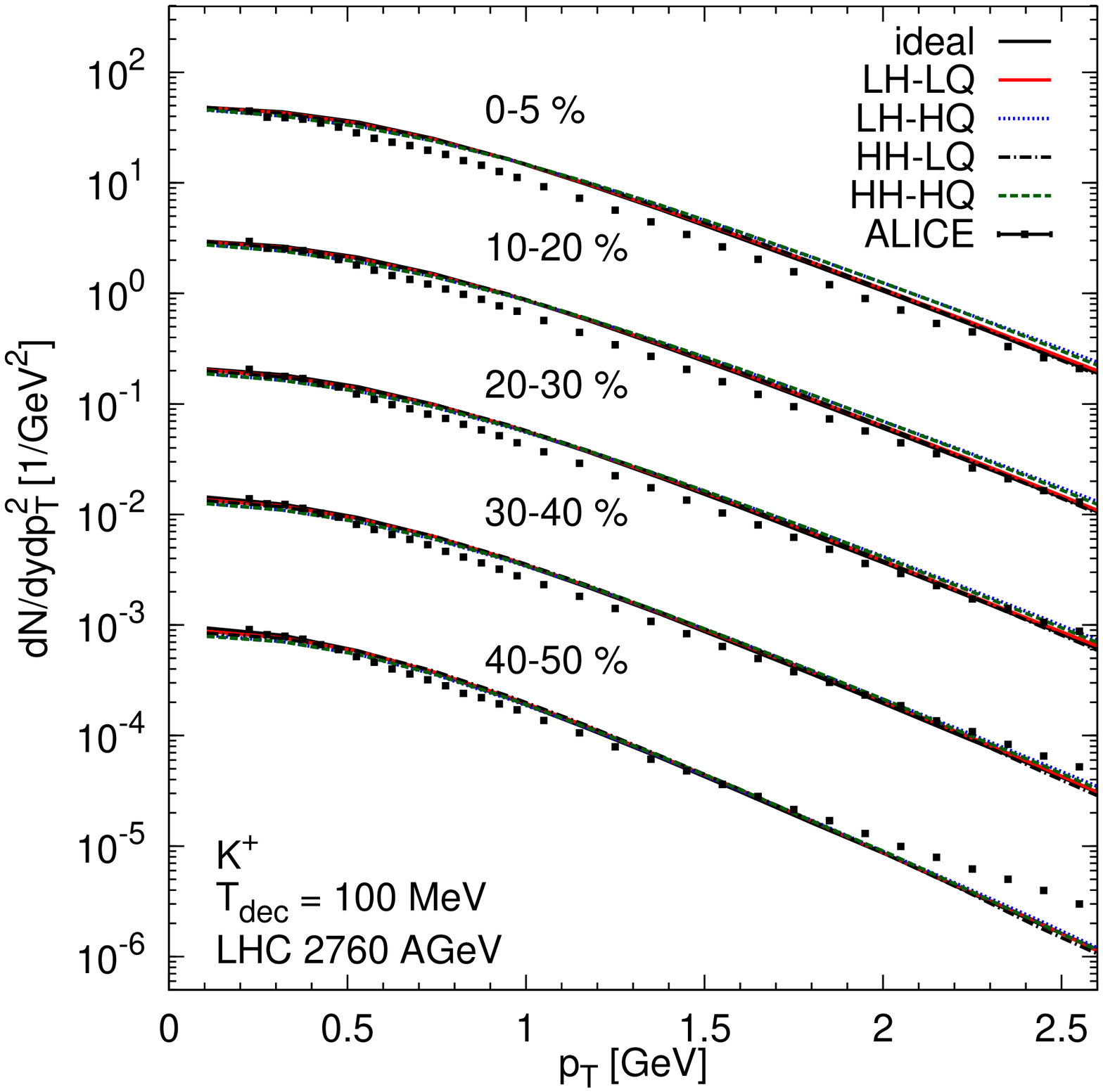} 
\vspace{-0.3cm}
\caption{{\protect\small (Color online) Transverse momentum spectra of
    positive kaons at the LHC. Experimental data are from the ALICE
    Collaboration \protect\cite{Abelev:2013vea}. }}
\label{fig:kaon_spectra_LHC}
\end{figure}
\begin{figure}
\vspace{-0.1cm} 
\includegraphics[width=8.5cm]{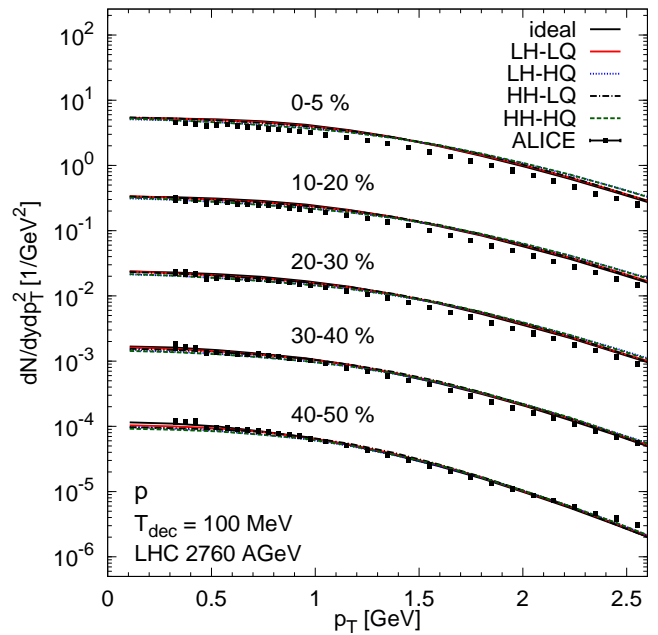} 
\vspace{-0.3cm}
\caption{{\protect\small (Color online) Transverse momentum spectra of
    protons at ALICE. Experimental data for protons and antiprotons
    are from the ALICE Collaboration \protect\cite{Abelev:2013vea}. }}
\label{fig:proton_spectra_LHC}
\end{figure}

In Figs.~{\ref{fig:pion_spectra_LHC}}, {\ref{fig:kaon_spectra_LHC}},
and~{\ref{fig:proton_spectra_LHC}} we show the $p_T$ spectra of
positive pions, positive kaons, and protons corresponding to
centrality classes, with multiplicative factors applied for better
visibility. The experimental data are from the ALICE
Collaboration~\cite{Abelev:2013vea}. These distributions behave in a
way similar to that of the RHIC results, and are thus unaffected by
the different $\eta_s/s$ parametrizations. We note that, as in many
other calculations~\cite{Abelev:2013vea,Begun:2013nga}, the low-$p_T$
part of the pion distribution turned out to be very difficult to
reproduce.

\begin{figure}
\vspace{-0.1cm} 
\includegraphics[width=8.5cm]{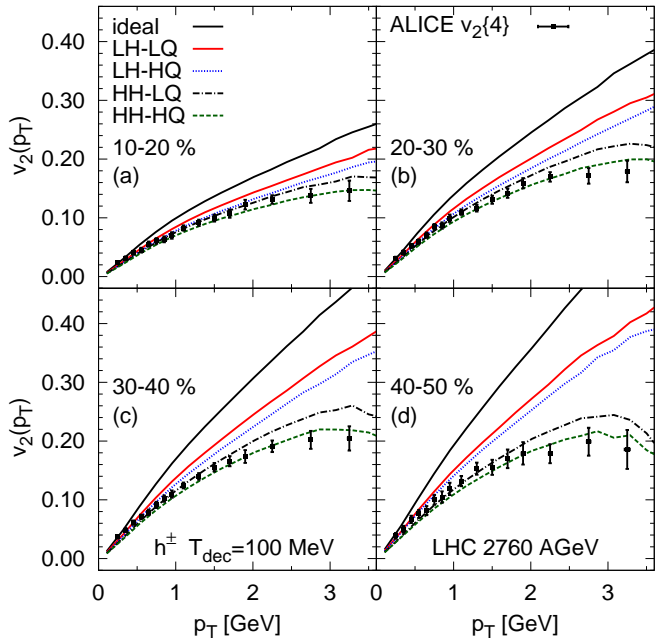} 
\vspace{-0.3cm}
\caption{{\protect\small (Color online) Charged hadron $v_2(p_T)$ at
    the LHC. Experimental data are from the ALICE Collaboration
    \cite{Aamodt:2010pa}. }}
\label{fig:charged_hadron_v2_LHC}
\end{figure}

\begin{figure}
\vspace{-0.1cm} 
\includegraphics[width=8.5cm]{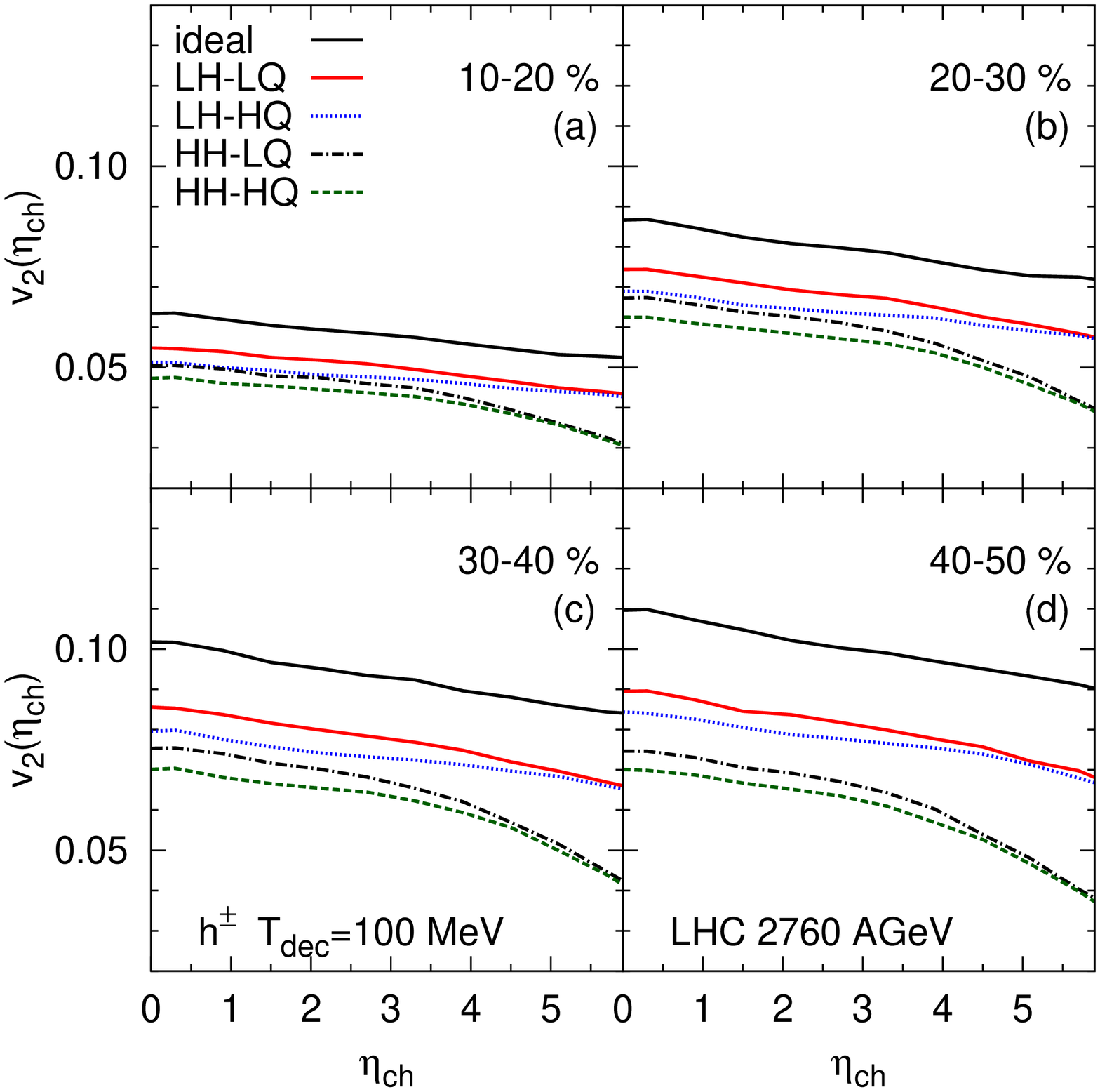} 
\vspace{-0.3cm}
\caption{{\protect\small (Color online) Charged hadron
    $v_2(\protect\eta_{ch})$ at the LHC. }}
\label{fig:charged_hadron_v2_y_LHC}
\end{figure}

In Figs.~\ref{fig:charged_hadron_v2_LHC}
and~\ref{fig:charged_hadron_v2_y_LHC} the elliptic flow coefficient
$v_2$ is shown as functions of transverse momentum and pseudorapidity,
respectively. In both figures the experimental data are from the ALICE
Collaboration~\cite{Aamodt:2010pa}. At the LHC viscous suppression of
the elliptic flow is less dominated by the hadronic viscosity than at
RHIC. In central collisions at midrapidity, both QGP and hadronic
viscosities affect $v_2$ equally: Large QGP viscosity may be
compensated with a low hadronic viscosity and vice versa (compare
LH-HQ with HH-LQ for 10--20\% and 20--30\% up to $p_T \leq 2$ GeV or
$\eta_{ch} \leq 2$). In peripheral collisions and at large rapidities
$v_2$ loses its sensitivity to QGP viscosity, and the system behaves
like at RHIC. Thus measuring $v_2$ at large rapidities at the LHC
would provide an additional handle on the temperature dependence of
the $\eta_s/s$ ratio.


\begin{figure}
\vspace{-0.1cm} 
\includegraphics[width=8.5cm]{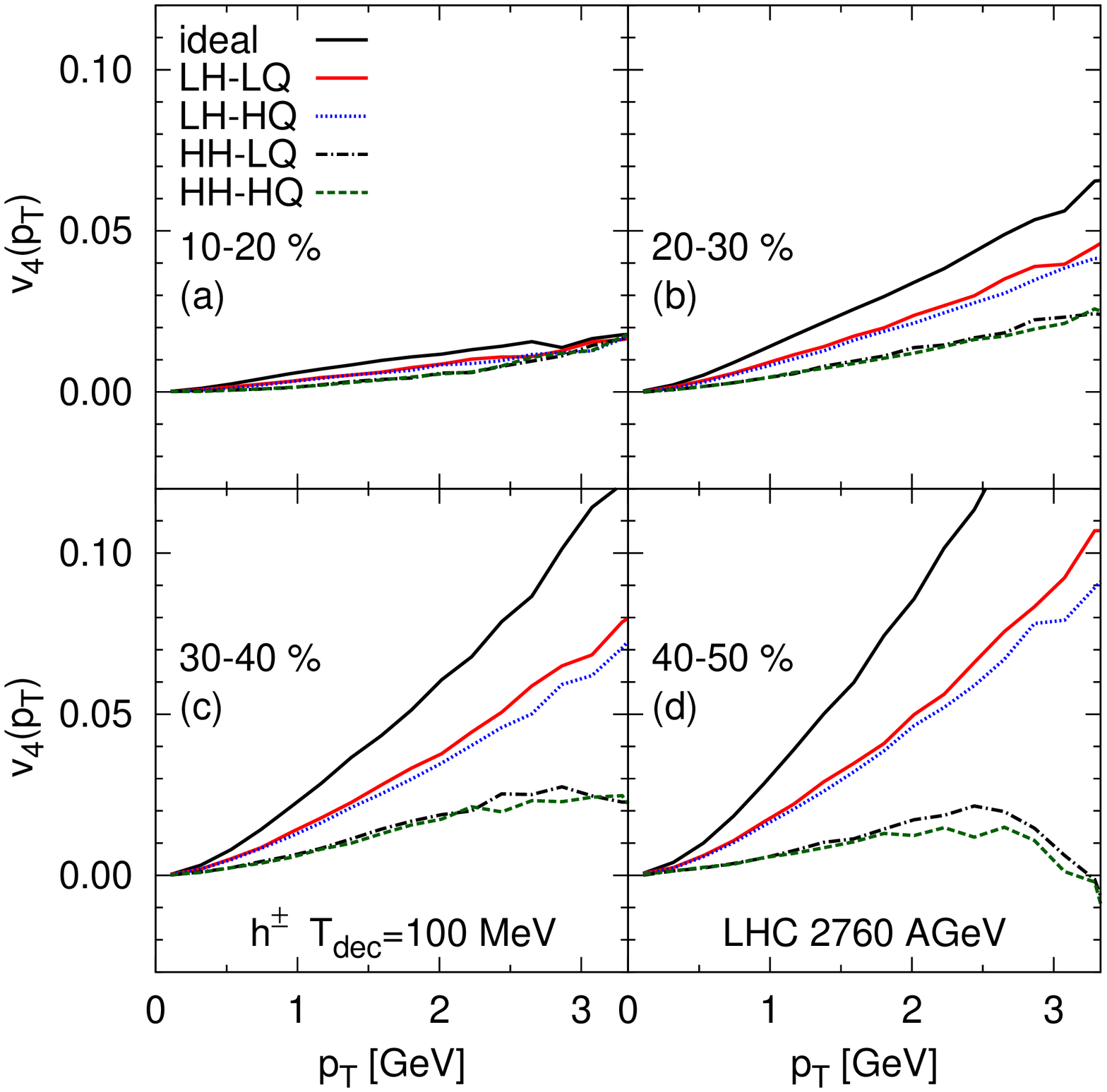} 
\vspace{-0.3cm}
\caption{{\protect\small (Color online) Charged hadron $v_4(p_{T})$ at
    the LHC.}}
\label{fig:charged_hadron_v4_LHC}
\end{figure}
\begin{figure}
\vspace{-0.1cm} 
\includegraphics[width=8.5cm]{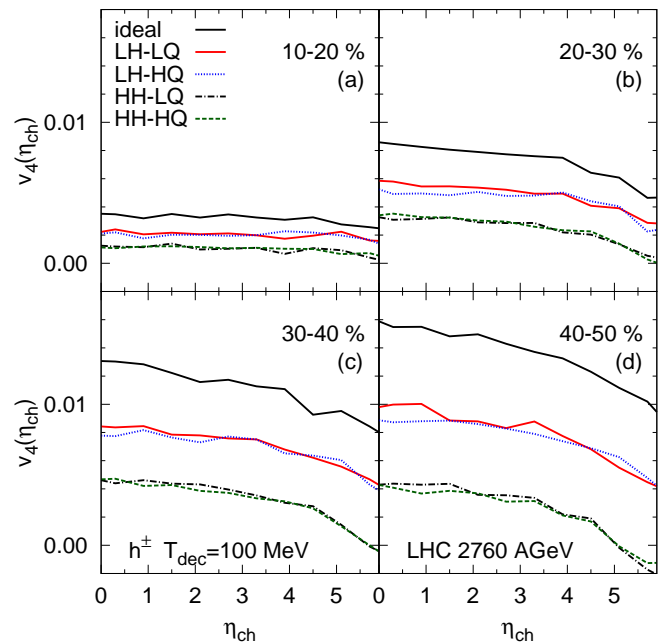} 
\vspace{-0.3cm}
\caption{{\protect\small (Color online) Charged hadron
    $v_4(\eta_{ch})$ at the LHC.}}
\label{fig:charged_hadron_v4_y_LHC}
\end{figure}

Finally, in Figs.~\ref{fig:charged_hadron_v4_LHC}
and~\ref{fig:charged_hadron_v4_y_LHC} we present the $v_4$
coefficients as functions of $p_T$ and $\eta_{ch}$. As discussed in
Refs.~\cite{SQM,Niemi:2012ry}, $v_4$ is sensitive to viscosity at
lower temperatures than $v_2$. Therefore the behavior of $v_4$ at the
LHC is similar to the behavior of $v_4$ and $v_2$ at RHIC: The curves
are grouped according to their hadronic viscosity, and show no
sensitivity to QGP viscosity. The suppression of $v_4$ at both the LHC
and RHIC is clearly sensitive to the hadronic viscosity (compare
Fig.~\ref{fig:charged_hadron_v4_RHIC} with
Fig.~\ref{fig:charged_hadron_v4_LHC} and
Fig.~\ref{fig:charged_hadron_v4_y_RHIC}
with~\ref{fig:charged_hadron_v4_y_LHC}) and to the minimum value of
$\eta_s/s$.

\section{The distinguishability of the $\eta_s/s$ parametrizations}
\label{rapidity_v2}

\begin{figure} 
\vspace{-0.1cm} %
\includegraphics[width=8cm]{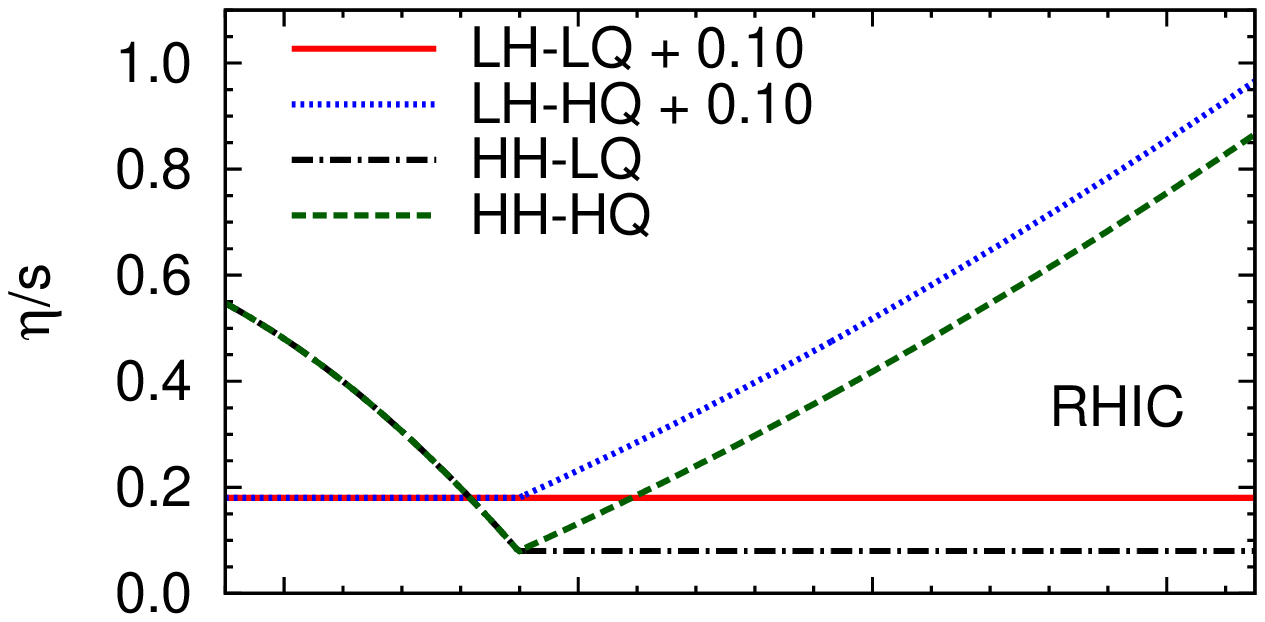} 
\includegraphics[width=8cm]{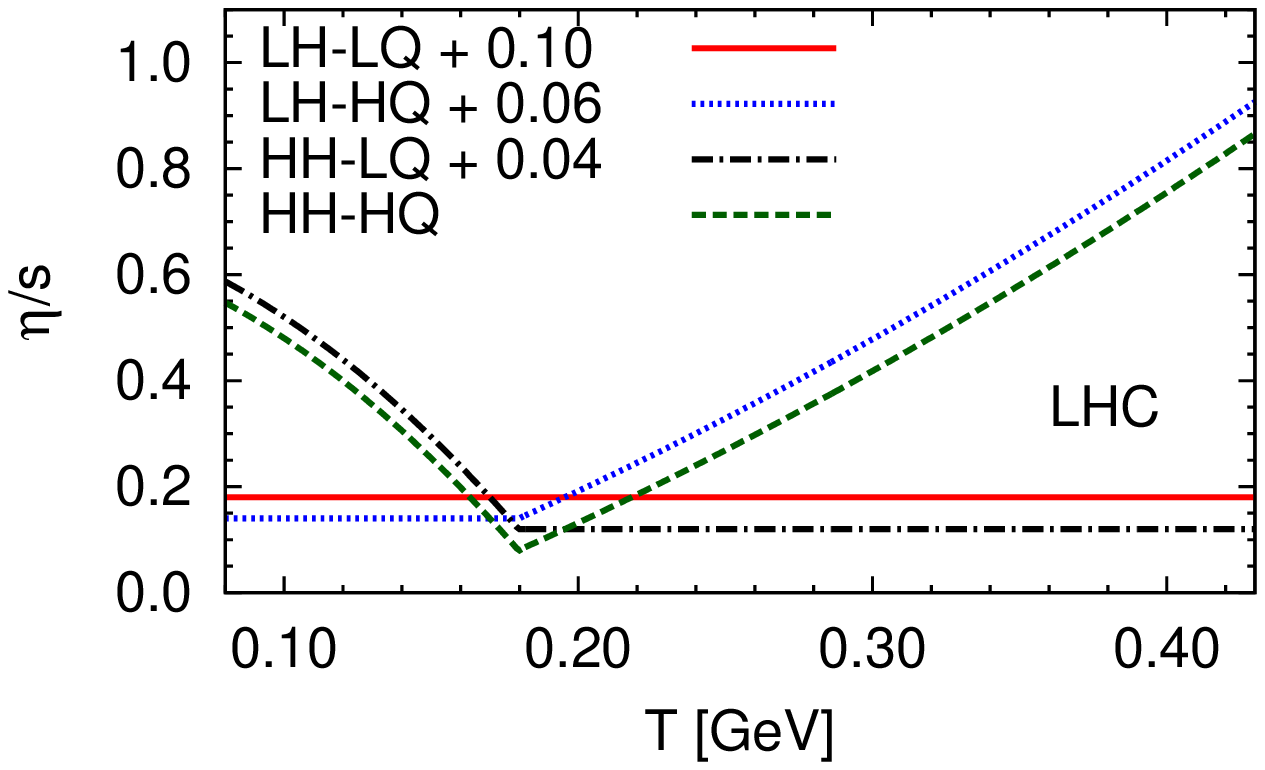}
\vspace{-0.3cm}
\caption{{\protect\small (Color online) Parametrizations of
    $(\eta_s/s)(T)$ rescaled to lead to similar charged hadron
    $v_2(p_T)$ in central collisions at RHIC (top) and the LHC (bottom).}}
\label{fig:etapers_scaled}
\end{figure}

In the previous section we described how the sensitivity of $v_2$ and
$v_4$ to QGP and hadronic shear viscosities depends on centrality,
transverse momentum $p_T$, and pseudorapidity $\eta_{ch}$. Now we use
this observation to distinguish between different parametrizations of
$(\eta_s/s)(T)$. We rescale our existing parametrizations in such a
way that they all lead to almost identical $p_T$ differential $v_2$ in
central collisions, and check whether the calculated $v_2$ and $v_4$
differ at other centralities and rapidities. Note that this procedure
also tests the sensitivity of the flow coefficients to the minimum
value of $\eta_s/s$, and not only to its values above and below the
transition temperature.

The new scaled parametrizations are shown in
Fig.~\ref{fig:etapers_scaled}. At RHIC energies the value of the
viscosity to entropy ratio for LH-LQ and LH-HQ is increased uniformly
with $\Delta \eta_s/s = 0.1$ for all temperatures, while the other two
parametrizations remain unchanged. Since the sensitivity to the
temperature dependence of $\eta_s/s$ is more complicated at the LHC,
the required changes in parametrizations are $\Delta \eta_s/s = 0.1$
for LH-LQ, $\Delta \eta_s/s = 0.06$ for LH-HQ and 
$\Delta\eta_s/s = 0.04$ for HH-LQ. The increase in $\eta_s/s$ leads to
larger entropy production, and thus to larger final multiplicities,
which we have counteracted by rescaling the initial densities
accordingly.

\begin{figure}
\vspace{-0.1cm} 
\includegraphics[width=8.5cm]{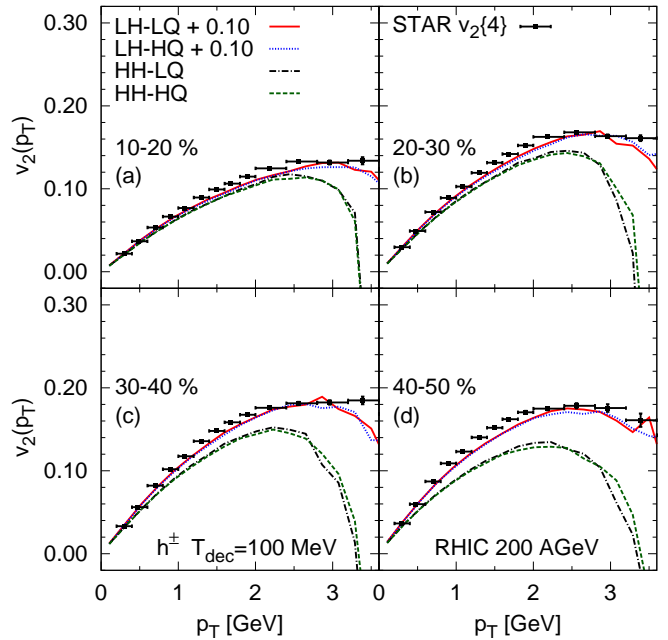} 
\vspace{-0.3cm}
\caption{{\protect\small (Color online) Charged hadron $v_2(p_T)$ at
    RHIC. Experimental data are from the STAR Collaboration
    \protect\cite{Bai}. }}
\label{fig:shift_charged_hadron_v2_RHIC}
\end{figure}
\begin{figure}
\vspace{-0.1cm} 
\includegraphics[width=8.5cm]{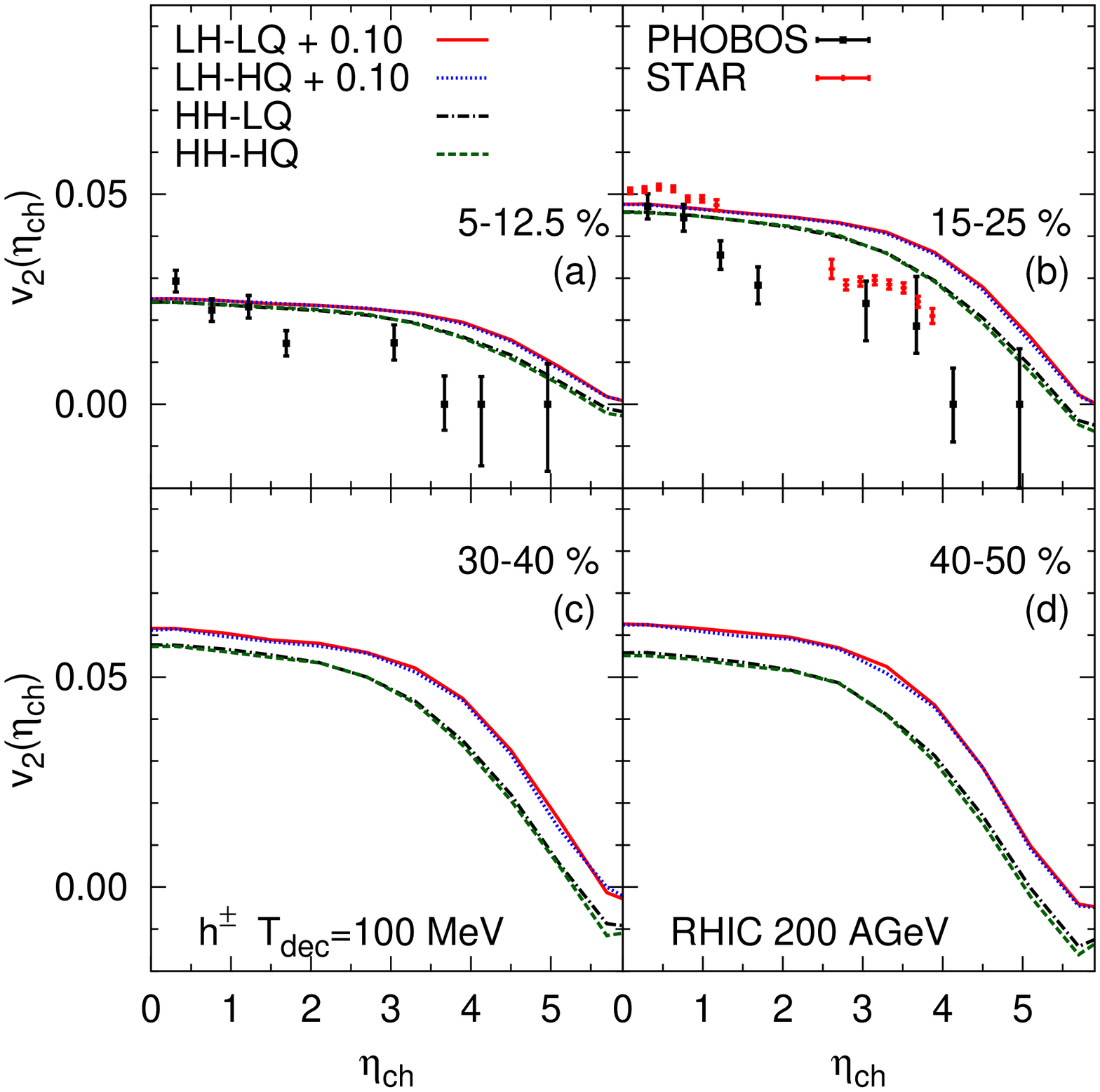}
\vspace{-0.3cm}
\caption{{\protect\small (Color online) Charged hadron
    $v_2(\protect\eta_{ch})$ at RHIC. Experimental data are from the
    PHOBOS \protect\cite{Back:2004mh} and STAR
    \protect\cite{Adams:2004bi} Collaborations. }}
\label{fig:shift_charged_hadron_v2_y_RHIC}
\end{figure}
\begin{figure}
\vspace{-0.1cm} 
\includegraphics[width=8.5cm]{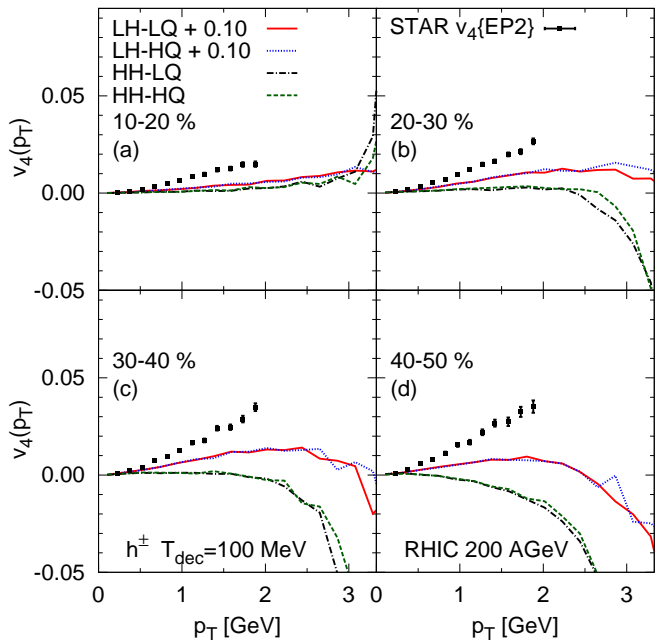} 
\vspace{-0.3cm}
\caption{{\protect\small (Color online) Charged hadron $v_4(p_T)$ at
    RHIC. Experimental data are from the STAR Collaboration
    \cite{Adams:2004bi}.}}
\label{fig:shift_charged_hadron_v4_RHIC}
\end{figure}

Note that since the LH-HQ and HH-LQ parametrizations require different
rescalings at RHIC and the LHC, they can be distinguished already by
comparing $v_2(p_T)$ in central collisions at different energies, but
LH-LQ and HH-HQ cannot. Furthermore, we want to check whether it is
possible to distinguish LH-HQ and HH-LQ in collisions at the same
energy by varying the centrality and rapidity.

In Figs.~\ref{fig:shift_charged_hadron_v2_RHIC},
\ref{fig:shift_charged_hadron_v2_y_RHIC},
and~\ref{fig:shift_charged_hadron_v4_RHIC} we present $v_2(p_T)$,
$v_2(\eta_{ch})$, and $v_4(p_T)$ at RHIC using these new
parametrizations. As required, in central collisions all
parametrizations lead to similar $v_2(p_T)$---the differences due to
different hadronic viscosity at very late stages of the evolution are
compensated by the larger viscosity at and after the QCD transition
region. However, when one moves to larger centralities, and thus to
smaller systems, the region where $v_2$ is most sensitive to shear
viscosity moves toward lower temperatures, and the parametrizations
with different hadronic viscosities can be identified, see
Fig.~\ref{fig:shift_charged_hadron_v2_RHIC}. The same, although
weaker, phenomenon happens when we move to larger rapidities, see
Fig.~\ref{fig:shift_charged_hadron_v2_y_RHIC}. Most of the sensitivity
comes from the change in centrality, but as seen in the 15--25\%
centrality class (Fig.~\ref{fig:shift_charged_hadron_v2_y_RHIC}b), the
difference at large rapidities increases faster than at
midrapidity. On the other hand, the $v_4$ coefficient shows larger
sensitivity than $v_2$: In central collisions all parametrizations are
equal, but the difference increases with increasing fraction of cross
section faster than for $v_2$. Note that none of the observables are
sensitive to the plasma viscosity, but we have to study the collisions
at the LHC to be able to distinguish, say, HH-LQ and HH-HQ
parametrizations.

\begin{figure}
\vspace{-0.1cm} 
\includegraphics[width=8.5cm]{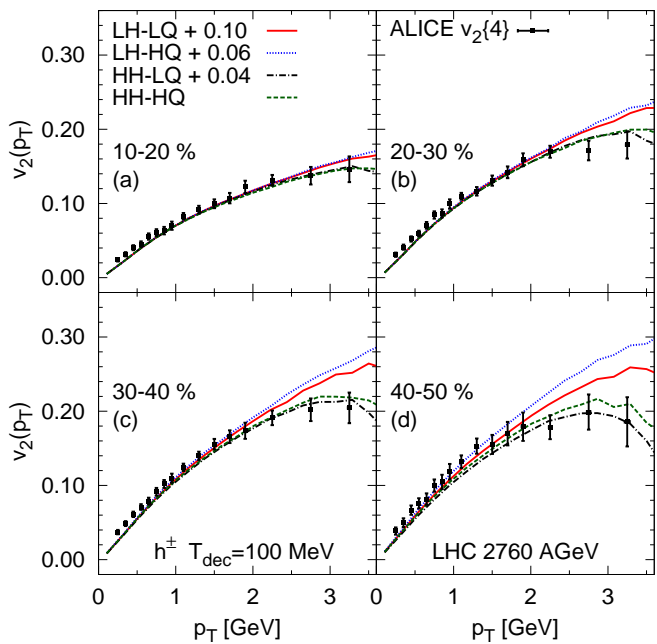} 
\vspace{-0.3cm}
\caption{{\protect\small (Color online) Charged hadron $v_2(p_T)$ at
    the LHC. Experimental data are from the ALICE Collaboration
    \cite{Aamodt:2010pa}. }}
\label{fig:shift_charged_hadron_v2_LHC}
\end{figure}
\begin{figure}
\vspace{-0.1cm} 
\includegraphics[width=8.5cm]{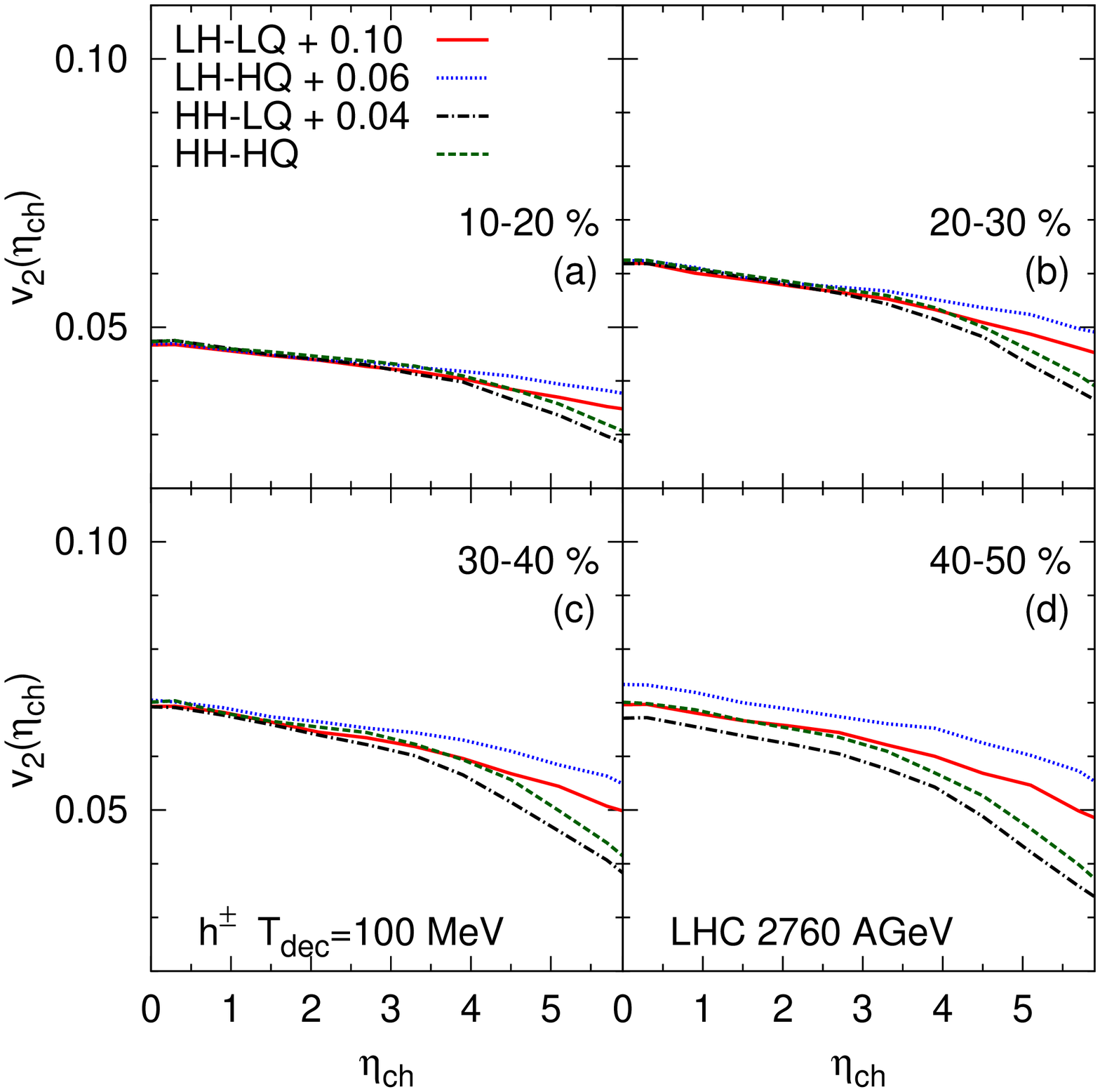} 
\vspace{-0.3cm}
\caption{{\protect\small (Color online) Charged hadron
    $v_2(\protect\eta_{ch})$ at the LHC. }}
\label{fig:shift_charged_hadron_v2_y_LHC}
\end{figure}
\begin{figure}
\vspace{-0.1cm} 
\includegraphics[width=8.5cm]{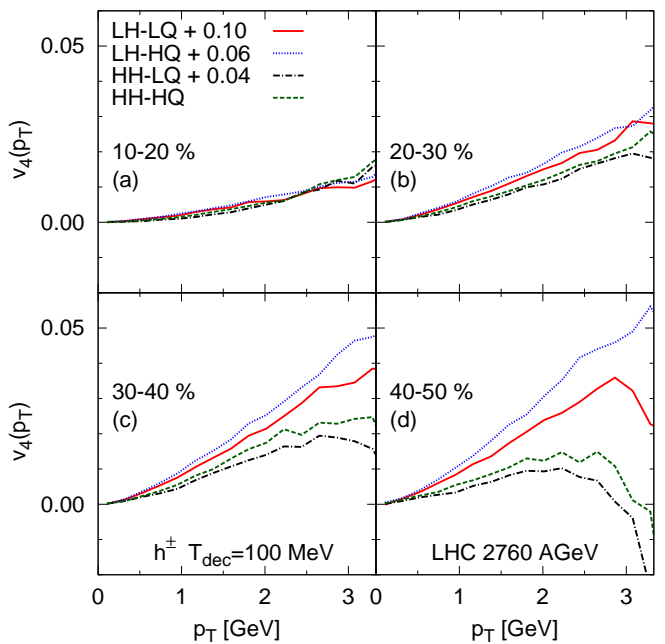} 
\vspace{-0.3cm}
\caption{{\protect\small (Color online) Charged hadron $v_4(p_{T})$ at
    the LHC.}}
\label{fig:shift_charged_hadron_v4_LHC}
\end{figure}

At the LHC we see slightly different behavior. In central collisions
$v_2(p_T)$ is again the same for all parametrizations by construction,
but the differences appear slowly and stay modest when we move toward
more peripheral collisions, see
Fig.~\ref{fig:shift_charged_hadron_v2_LHC}. Again, in more peripheral
collisions, the system is most sensitive to viscosity in lower
temperatures, and $v_2(p_T)$ curves are ordered according to hadronic
viscosity---the larger viscosity at freeze-out, the lower $v_2(p_T)$.
In Fig.~\ref{fig:charged_hadron_v2_y_LHC} the pseudorapidity
distribution of $v_2$ showed clear sensitivity to shear viscosity. In
that figure different parametrizations caused different $v_2$ already
at midrapidity in central collisions. Now viscosity is scaled to
remove this difference, and the sensitivity of the shape of
$v_2(\eta_{ch})$ to the viscosity is more visible. As one can see from
Fig.~\ref{fig:shift_charged_hadron_v2_y_LHC}, larger hadronic
viscosity causes $v_2(\eta_{ch})$ to drop slightly faster with
increasing rapidity. The strongest difference is seen in $v_4(p_T)$,
which is able to distinguish the new parametrizations at the LHC, see
Fig.~\ref{fig:shift_charged_hadron_v4_LHC}, but its resolving power at
the LHC is weaker than at RHIC
(Fig.~\ref{fig:shift_charged_hadron_v4_RHIC}). Thus we conclude that
differential measurements of the flow anisotropies as function of
transverse momentum, pseudorapidity, and centrality can provide
constraints for the temperature dependence of $\eta_s/s$, but the
measurements at various energies are essential to constrain the
parametrizations properly.

\section{Dynamical freeze-out}
\label{dynamical_FO}

\begin{figure}
\vspace{-0.1cm} 
\includegraphics[width=8.5cm]{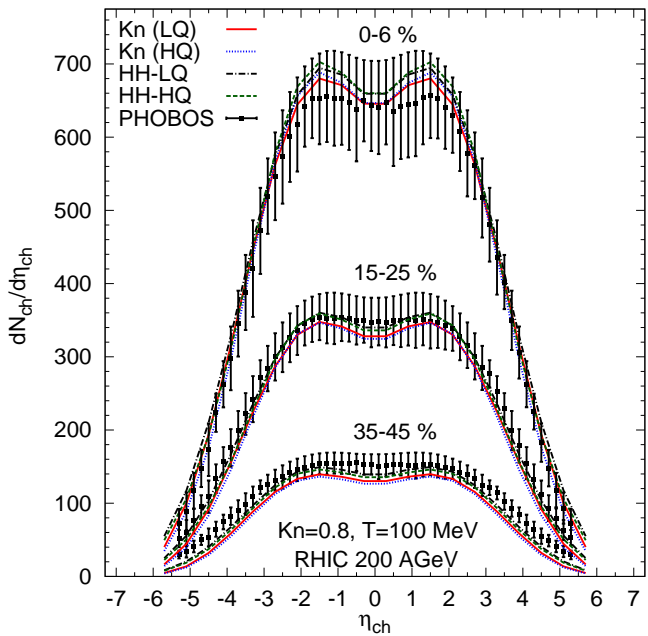} 
\vspace{-0.3cm}
\caption{{\protect\small (Color online) The charged particle
    pseudorapidity distribution $dN_{ch}/d\protect\eta_{ch}$ at RHIC
    obtained by using two different freeze-out criteria. Experimental
    data are from the PHOBOS Collaboration
    \protect\cite{Back:2002wb}.}}
\label{fig:charged_hadron_y_RHIC_Kn}
\end{figure}
\begin{figure}
\vspace{-0.1cm} 
\includegraphics[width=8.5cm]{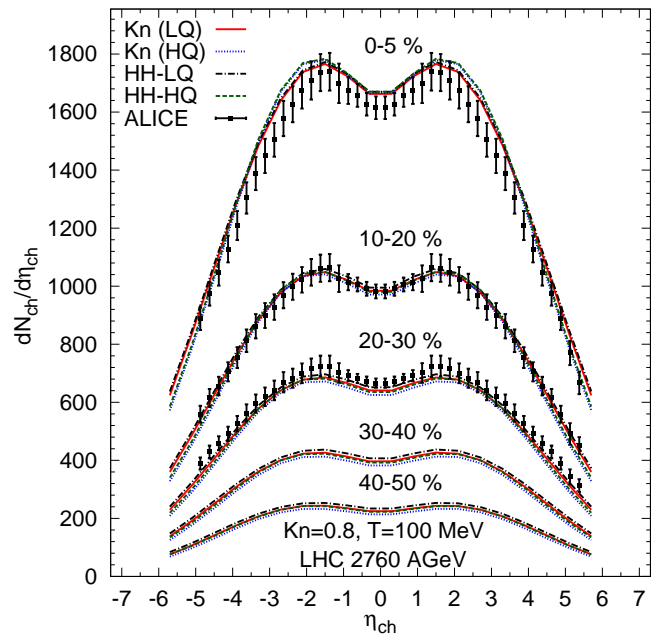} 
\vspace{-0.3cm}
\caption{{\protect\small (Color online) The charged particle
    pseudorapidity distribution $dN_{ch}/d\protect\eta_{ch}$ at the
    LHC obtained by using two different freeze-out
    criteria. Experimental data are from the ALICE Collaboration
    \protect\cite{Abbas:2013bpa}. }}
\label{fig:charged_hadron_y_LHC_Kn}
\end{figure}

To test the sensitivity of our results to the freeze-out criterion and
the freeze-out description in general, we redo some of the
calculations using the dynamical freeze-out
criterion~\cite{Bondorf:1978kz}. In these calculations we use only our
HH-LQ and HH-HQ parametrizations for the shear viscosity, since the
low value of $\eta_s/s$ in a hadron gas leads to a very slowly
increasing relaxation time and thus to unrealistically low
temperatures, $\langle T \rangle \ll 80$ MeV on the freeze-out surface
when $\text{Kn}_{dec}\sim 1$.  Since the Knudsen number can be based
on many quantities~\cite{Niemi:2014wta}, and since we do not know when
exactly the hydrodynamical description should break down, we use the
freeze-out Knudsen number as a free parameter chosen to fit the
rapidity and $p_T$ distributions.

Figures~\ref{fig:charged_hadron_y_RHIC_Kn}
and~\ref{fig:charged_hadron_y_LHC_Kn} show the charged particle
pseudorapidity distributions at RHIC and the LHC, respectively. As
expected, the pseudorapidity distributions are only weakly dependent
on the precise value of $\text{Kn}_{dec}$, but it turned out that our
choice of Knudsen number and relaxation time lead to weak sensitivity
of the $p_T$ distributions to the value of $\text{Kn}_{dec}$
too. Nevertheless, we found that decoupling at constant Knudsen number
$\text{Kn}_{dec} = 0.8$ leads to basically the same rapidity and $p_T$
distributions as the conventional decoupling at $T_{dec} = 100$ MeV.

\begin{figure}
\vspace{-0.1cm} 
\includegraphics[width=8.5cm]{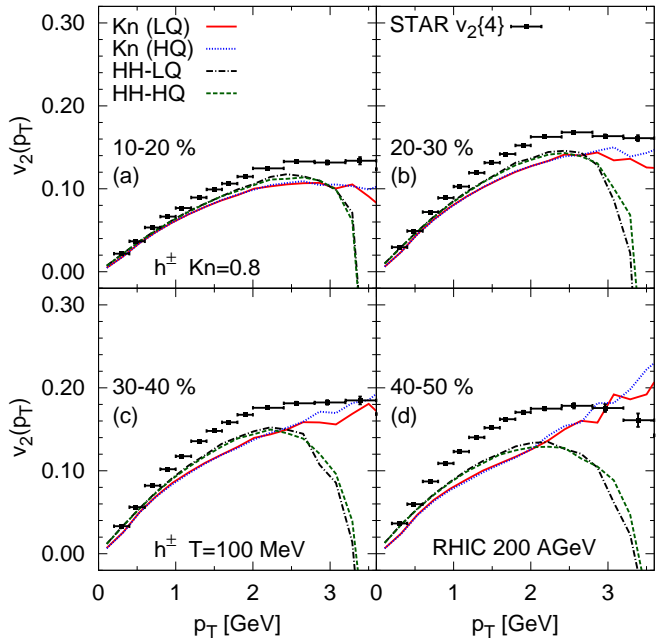} 
\vspace{-0.3cm}
\caption{{\protect\small (Color online) Charged hadron $v_2(p_T)$ at
    RHIC obtained by using two different freeze-out
    criteria. Experimental data are from the STAR Collaboration
    \protect\cite{Bai}. }}
\label{fig:charged_hadron_v2_RHIC_Kn}
\end{figure}

\begin{figure}
\vspace{-0.1cm} 
\includegraphics[width=8.5cm]{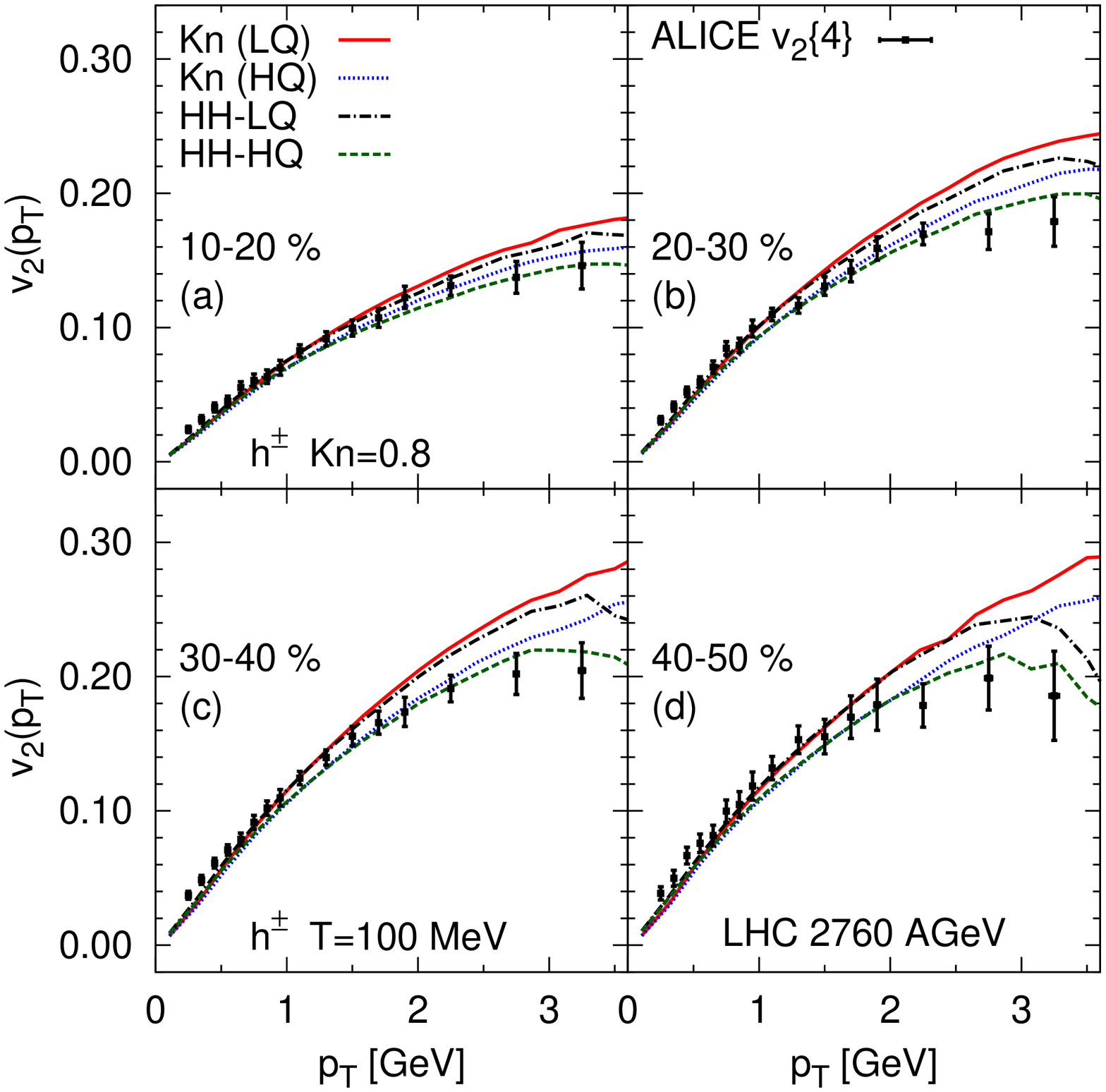} 
\vspace{-0.3cm}
\caption{{\protect\small (Color online) Charged hadron $v_2(p_T)$ at
    the LHC obtained by using two different freeze-out
    criteria. Experimental data are from the ALICE Collaboration
    \cite{Aamodt:2010pa}. }}
\label{fig:charged_hadron_v2_LHC_Kn}
\end{figure}

The $p_T$ differential $v_2$ of charged hadrons at RHIC and the LHC is
shown in Figs.~\ref{fig:charged_hadron_v2_RHIC_Kn}
and~\ref{fig:charged_hadron_v2_LHC_Kn}, respectively. Unlike in
Ref.~\cite{Holopainen:2013jna}, where both $p_T$ distributions and
anisotropies depended on the freeze-out criterion, we see that once
the freeze-out parameters are fixed to produce similar $p_T$
distributions, the anisotropies become very similar. This is
especially clear at the LHC. Below $p_T \sim 2$ GeV both criteria lead
to identical $v_2(p_T)$, and the difference seen in the plots is due
to the shear viscosity parametrization. At RHIC both parametrizations
lead to identical $v_2(p_T)$, and a weak sensitivity to the freeze-out
criterion appears around $p_T\sim 1$ GeV. However, this sensitivity is
too weak to be significant.

\begin{figure} 
\vspace{-0.1cm} 
\includegraphics[width=8.5cm]{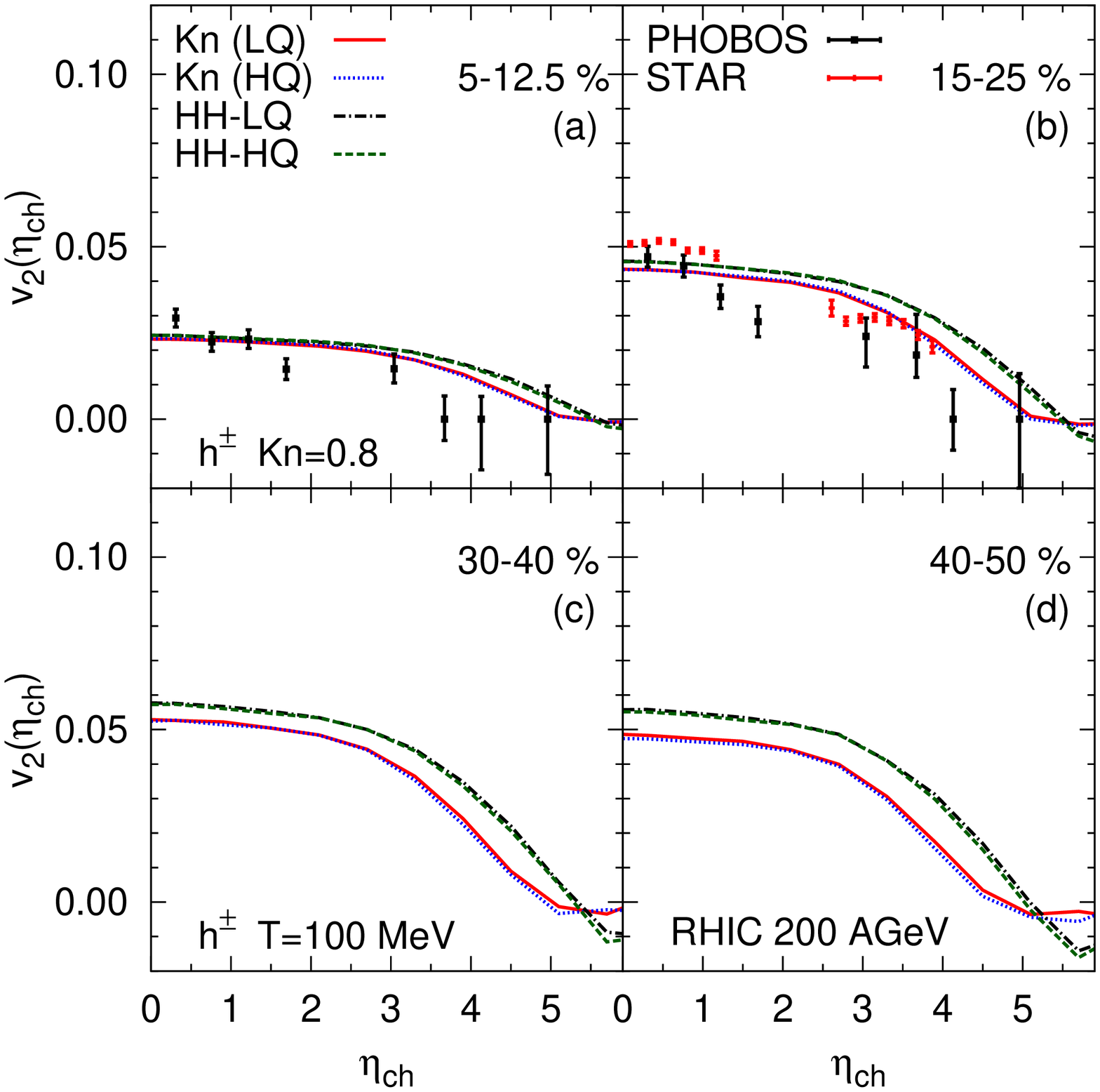} 
\vspace{-0.3cm}
\caption{{\protect\small (Color online) Charged hadron
    $v_2(\eta_{ch})$ at RHIC obtained by using two different freeze-out
    criteria. Experimental data are from the STAR Collaboration
    \protect\cite{Bai}. }}
\label{fig:charged_hadron_v2_y_RHIC_Kn}
\end{figure}
\begin{figure} 
\vspace{-0.1cm} 
\includegraphics[width=8.5cm]{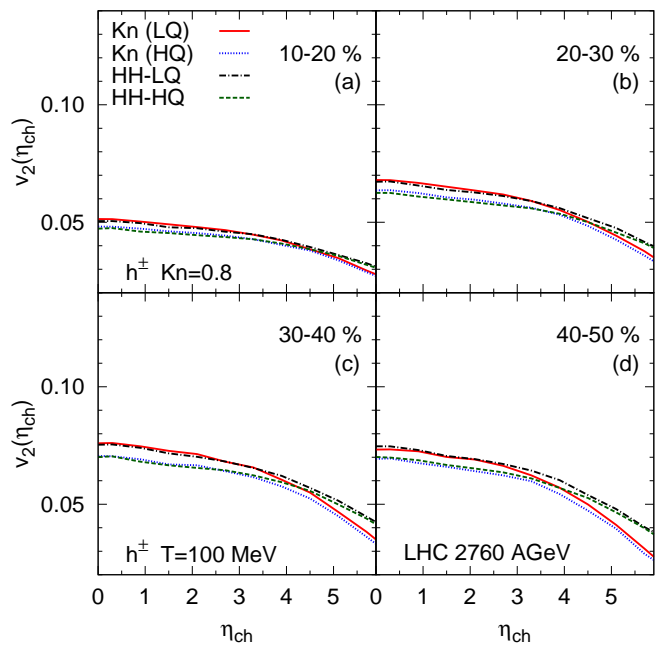} 
\vspace{-0.3cm}
\caption{{\protect\small (Color online) Charged hadron
    $v_2(\eta_{ch})$ at the LHC obtained by using two different freeze-out
    criteria. Experimental data are from the ALICE Collaboration
    \cite{Aamodt:2010pa}. }}
\label{fig:charged_hadron_v2_y_LHC_Kn}
\end{figure}

As a function of pseudorapidity $v_2$ shows more sensitivity to the
freeze-out criterion, see Figs.~\ref{fig:charged_hadron_v2_y_RHIC_Kn}
and~\ref{fig:charged_hadron_v2_y_LHC_Kn}. At both RHIC and the LHC
$v_2(\eta_{ch})$ drops faster with increasing rapidity, when the
dynamical freeze-out criterion is used. Also, with both freeze-out
criteria the sensitivity to plasma viscosity disappears at large
rapidities even at the LHC. This is again a manifestation of
previously seen behavior: At large rapidities at the LHC, the system
behaves like the system at RHIC.

The rather weak dependence of anisotropies on the decoupling criterion
means that at midrapidity fluid-dynamical results are surprisingly
robust against variations in the decoupling procedure. As well, this
gives a reason to expect that the hybrid model results are sensitive
only to the value of the switching criterion from fluid to cascade,
not to the criterion itself. Since the fluid-dynamical results
concerning the viscosity of the QGP are based on the analysis of
anisotropies at midrapidity, this means that those results are not
compromised by the freeze-out criterion. On the other hand, the
sensitivity to the freeze-out description at high rapidities indicates
that at lower collision energies the fluid-dynamical results may be
sensitive to the freeze-out criterion even at midrapidity. Thus one
has to pay extra attention to the freeze-out description of the
collisions at $\sqrt{s_{NN}} = 3-9$ GeV in the future FAIR and NICA
facilities.

\section{Conclusions}
\label{conclusions}

We have studied the effects of temperature dependent $\eta_s/s$ on the
azimuthal anisotropies of hadron transverse momentum spectra using
genuinely (3+1)-dimensional viscous hydrodynamics. We have extended
our previous studies~\cite{Niemi:2011ix,Niemi:2012ry} to back- and
forward rapidities and explored the resolving power of differential
measurements of $v_2$ and $v_4$ to distinguish between different
parametrizations of $(\eta_s/s)(T)$.

In close to central collisions at the LHC energy,
$\sqrt{s_\mathrm{NN}} = 2.76$ TeV, viscous suppression of elliptic
flow at midrapidity is affected by both hadronic and QGP viscosities,
but when one moves toward back- and forward rapidities, hadronic
viscosity becomes more and more dominant---the system becomes
effectively smaller, and begins to behave like in collisions at RHIC,
$\sqrt{s_\mathrm{NN}} = 200$ GeV. Therefore with large hadronic
viscosity $v_2$ tends to drop slightly faster with increasing
rapidity, the effect being stronger in peripheral collisions. At both
energies and at all rapidities $v_4$ is mostly suppressed by hadronic
viscosity, but if we simultaneously change the minimum value of
$\eta_s/s$, hadronic, and QGP viscosities, it is difficult to predict
which coefficient at which collision energy is most sensitive to the
changes. Nevertheless, the differential measurements of $v_n$ as
functions of transverse momentum, rapidity, centrality, and collision
energy provide a way to distinguish different parametrizations of
$(\eta_s/s)(T)$, and thus constrain the temperature dependence of the
$\eta_s/s$ ratio.

We also studied how sensitive our results are to the freeze-out
criterion, and found that once the freeze-out parameters are fixed to
reproduce $p_T$ distributions, both decoupling at constant temperature
and at constant Knudsen number lead to very similar anisotropies at
midrapidity. Toward the large rapidities $v_2$ tends to drop faster
with the dynamical freeze-out criterion. This indicates that
uncertainties in the decoupling description do not affect the present
fluid-dynamical results regarding the anisotropies, but at lower
collision energies the results may be more sensitive to the freeze-out
criterion.

\begin{acknowledgments}
  This work was supported by the Helmholtz International Center for
  FAIR within the framework of the LOEWE program launched by the State
  of Hesse. The work of H.\ Niemi was supported by Academy of Finland,
  Project No.~133005, the work of P.~Huovinen by BMBF under Contract
  No.~06FY9092 and the work of H.~Holopainen by the ExtreMe Matter
  Institute (EMMI). E.\ Moln\'ar was partially supported by the
  European Union and the European Social Fund through project
  Supercomputer, the national virtual laboratory (Grant
  No.:\ TAMOP-4.2.2.C-11/1/KONV-2012-0010), as well as by TAMOP
  4.2.4.~A/2-11-1-2012-0001 National Excellence Program
  (A2-MZPD\"O-13-0042).
\end{acknowledgments}

\appendix

\section{Equations in (3+1)--dimensions}

\label{3+1d_equations}

In the following the components of four-vectors and tensors of rank-2
in four-dimensional space-time are denoted by Greek indices that take
values from $0$ to $3$ while Roman indices range from $1$ to $3$. If
not stated otherwise the Einstein summation convention for both Greek
and Roman indices is implied.

First we recall the definitions of the covariant derivative of
contravariant four-vectors and tensors of rank-2: 
\begin{align}
 A_{;\alpha}^\mu     & = \partial_{\alpha}A^\mu + \Gamma_{\alpha\beta}^{\mu}A^{\beta}\,, \\
 A_{;\alpha}^{\mu\nu} & = \partial_{\alpha}A^{\mu\nu}
                    + \Gamma_{\alpha\beta}^{\mu}A^{\beta\nu}
                    + \Gamma_{\alpha\beta}^{\nu }A^{\mu \beta }\,,
\end{align}
where $\Gamma_{\alpha\beta}^{\mu}\equiv \Gamma_{\beta \alpha }^{\mu} =
\frac{1}{2}\,g^{\mu \nu }\left( \partial _{\beta }g_{\alpha \nu }+\partial
_{\alpha }g_{\nu \beta }-\partial _{\nu }g_{\alpha \beta }\right)$ denotes
the Christoffel symbol of the second kind and $\partial _{\alpha }=\partial
/\partial x^{\alpha }$ denotes the four-derivative. For scalar quantities
the covariant derivative reduces to the ordinary four-derivative, i.e.,
$\left(A^{\mu}A_{\mu}\right)_{;\alpha} = \partial_{\alpha}\left(A^{\mu}A_{\mu}\right) $. 

Applying the definition of the transverse projection operator
$\Delta ^{\mu\nu } = g^{\mu \nu }-u^{\mu }u^{\nu }$ we can decompose the covariant
derivative as the sum of the covariant time derivative $D$ and spatial
gradient $\nabla _{\alpha }$, 
\begin{eqnarray} \label{D_derivative}
DA^{\mu _{1}\cdots \mu _{n}} &=& u^{\beta }A_{;\beta }^{\mu _{1}\cdots \mu
_{n}}, \\
\nabla _{\alpha }A^{\mu _{1}\cdots \mu _{n}} &=&\Delta _{\alpha }^{\beta
}A_{;\beta }^{\mu _{1}\cdots \mu _{n}};
\end{eqnarray}%
hence $A_{;\alpha }^{\mu _{1}\cdots \mu _{n}}=u_{\alpha }DA^{\mu _{1}\cdots
\mu _{n}}+\nabla _{\alpha }A^{\mu _{1}\cdots \mu _{n}}$, while for later use
we also introduce the comoving or convective time derivative 
\begin{equation}\label{d_derivative}
dA^{\mu _{1}\cdots \mu _{n}}=u^{\beta }\partial _{\beta }A^{\mu _{1}\cdots
\mu _{n}}.
\end{equation}

In the following we summarize the equations of relativistic
dissipative fluid dynamics in hyperbolic coordinates [\emph{i.e.}\
$\left( \tau,x,y,\eta \right)$--coordinates]~\cite{Bjorken:1982qr}, where
$\tau=(t^{2}-z^{2})^{-1/2}$ is the longitudinal proper time and
$\eta = 1/2\ln \left[ (t+z)/(t-z)\right] $ is the space-time
rapidity. The proper metric tensors are
$g^{\mu\nu} = \text{diag}(1,-1,-1,-\tau^{-2})$ and $g_{\mu \nu
}=\text{diag}(1,-1,-1,-\tau ^{2})$. Thus the only nonvanishing
Christoffel symbols are $\Gamma _{\eta \tau }^{\eta }\equiv \Gamma
_{\tau \eta }^{\eta }=\tau ^{-1}$ and $\Gamma _{\eta \eta }^{\tau
}=\tau $, and the gradient is $\partial _{\mu }=(\partial _{\tau
},\partial _{x},\partial _{y},\partial _{\eta })$ while $\partial
^{\mu }\equiv g^{\mu \nu }\partial _{\nu }=(\partial _{\tau
},-\partial _{x},-\partial _{y},-\tau ^{-2}\partial _{\eta })$. The
inverse transformations to Minkowski coordinates with $%
g^{\mu \nu }_M\equiv \eta ^{\mu \nu }=\text{diag}(1,-1,-1,-1) $ are
$t=\tau \cosh \eta $ and $z=\tau \sinh \eta $. Note that the hyperbolic
coordinates are similar to the Milne coordinates that are spherically
symmetric, i.e., $%
r\equiv \sqrt{x^{2}+y^{2}+z^{2}}=\tau \sinh \eta $.

The contravariant flow velocity is 
\begin{equation}
u^{\mu }=\gamma\left( 1,v_{x},v_{y},v_{\eta }\right); \label{contravariant_u}
\end{equation} 
hence the covariant flow velocity is 
$u_{\mu }\equiv g_{\mu \nu }u^{\nu }=\gamma \left( 1,-v_{x},-v_{y},-\tau
^{2}v_{\eta }\right) $, where the normalization condition $u^{\mu} u_{\nu}=1$
leads to $\gamma
=\left( 1-v_{x}^{2}-v_{y}^{2}-\tau ^{2}v_{\eta }^{2}\right) ^{-1/2}$ as well as 
to $u^{\mu }u_{\mu ;\nu }\equiv
u_{\mu }u_{;\nu }^{\mu }=0$.

The energy-momentum conservation equation in general coordinates is
\begin{equation}
T_{;\mu }^{\mu \nu }\equiv \frac{1}{\sqrt{g}}\partial _{\mu }\left( \sqrt{g}%
\,T^{\mu \nu }\right) +\Gamma _{\mu \beta }^{\nu }T^{\mu \beta }=0,
\label{em_cons_1}
\end{equation}%
where $g\equiv -\text{det}(g_{\mu \nu })$ is the negative determinant of the
metric tensor, which in hyperbolic coordinates leads to $g=\tau ^{2}$.

Henceforth the energy conservation equation leads to
\begin{align}
\lefteqn{\partial _{\tau }T^{\tau \tau }+\partial _{x}(v_{x}T^{\tau \tau
})+\partial _{y}(v_{y}T^{\tau \tau })+\partial _{\eta }(v_{\eta }T^{\tau
\tau })}  \notag \\
& =-\partial _{x}\left( v_{x}P-v_{x}\pi ^{\tau \tau }+\pi ^{\tau x}\right)
-\partial _{y}\left( v_{y}P-v_{y}\pi ^{\tau \tau }+\pi ^{\tau y}\right) 
\notag \\
& -\partial _{\eta }\left( v_{\eta }P-v_{\eta }\pi ^{\tau \tau }+\pi ^{\tau
\eta }\right) -\frac{1}{\tau }\left( T^{\tau \tau }+\tau ^{2}T^{\eta \eta
}\right) \,,
\end{align}%
while the momentum-conservation equation leads to 
\begin{align}
\lefteqn{\partial _{\tau }T^{\tau x}+\partial _{x}(v_{x}T^{\tau x})+\partial
_{y}(v_{y}T^{\tau x})+\partial _{\eta }(v_{\eta }T^{\tau x})}  \notag \\
& =-\partial _{x}\left( P-v_{x}\pi ^{\tau x}+\pi ^{xx}\right) -\partial
_{y}\left( -v_{y}\pi ^{\tau x}+\pi ^{xy}\right)  \notag \\
& -\partial _{y}\left( -v_{\eta }\pi ^{\tau x}+\pi ^{x\eta }\right) -\frac{1%
}{\tau }T^{\tau x}\,, \\
\lefteqn{\partial _{\tau }T^{\tau y}+\partial _{x}(v_{x}T^{\tau y})+\partial
_{y}(v_{y}T^{\tau y})+\partial _{\eta }(v_{\eta }T^{\tau y})}  \notag \\
& =-\partial _{x}\left( -v_{x}\pi ^{\tau y}+\pi ^{xy}\right) -\partial
_{y}\left( P-v_{y}\pi ^{\tau y}+\pi ^{yy}\right)  \notag \\
& -\partial _{\eta }\left( -v_{\eta }\pi ^{\tau y}+\pi ^{y\eta }\right) -%
\frac{1}{\tau }T^{\tau y}\,, \\
\lefteqn{\partial _{\tau }T^{\tau \eta }+\partial _{x}(v_{x}T^{\tau \eta
})+\partial _{y}(v_{y}T^{\tau \eta })+\partial _{\eta }(v_{\eta }T^{\tau
\eta })}  \notag \\
& =-\partial _{x}\left( -v_{x}\pi ^{\tau \eta }+\pi ^{x\eta }\right)
-\partial _{y}\left( -v_{y}\pi ^{\tau \eta }+\pi ^{y\eta }\right)  \notag \\
& -\partial _{\eta }\left( \frac{P}{\tau ^{2}}-v_{\eta }\pi ^{\tau \eta
}+\pi ^{\eta \eta }\right) -\frac{3}{\tau }T^{\tau \eta }\,.
\end{align}
The corresponding tensor components are defined according to the general
definition of the energy-momentum tensor [Eq.~(\ref{T_munu_general})], 
\begin{align}
T^{\tau \tau }& = (e+P)\gamma ^{2}-g^{\tau \tau }P+\pi ^{\tau \tau }\ ,
\label{T^00} \\
T^{\tau i}& \equiv (e+P)\gamma ^{2}v_{i}-g^{\tau i}P+\pi ^{\tau i}\ ,  \notag
\\
& =v_{i}T^{\tau \tau }+P(g^{\tau \tau }v_{i}-g^{\tau i})-v_{i}\pi ^{\tau
\tau }+\pi ^{\tau i}\ ,  \label{T^0i} \\
T^{ij}& \equiv (e+P)\gamma ^{2}v_{i}v_{j}-Pg^{ij}+\pi ^{ij}\ ,  \notag \\
& =v_{i}T^{\tau j}+P(g^{\tau j}v_{i}-g^{ij})-v_{i}\pi ^{\tau j}+\pi ^{ij}\ .
\label{T^ij}
\end{align}

A simplified but mathematically equivalent way of writing the
equations of motion can be obtained by introducing scaled variables that
absorb the $\sqrt{g}$ factor~\cite{Kolb:2000sd},
\begin{equation}
\tilde{T}^{\mu \nu }=\tau T^{\mu \nu }\,; \label{scaling}
\end{equation}%
hence we are led to the following $\tau$--scaled equations: 
\begin{eqnarray}
{\partial }_{\mu }\tilde{T}^{\tau \mu }& =&-\tau\tilde{T}^{\eta \eta }\ , \\
{\partial }_{\mu }\tilde{T}^{x\mu }& =&0\ , \quad {\partial }_{\mu }\tilde{T}^{y\mu
}=0\ , \quad {\partial }_{\mu }\tilde{T}^{\eta \mu }=-\frac{2}{\tau}\tilde{T}%
^{\tau \eta }\ . \qquad
\end{eqnarray}
For example in special test cases with no transverse pressure and
vanishing dissipation we can solve the energy-conservation equation exactly.  
We found that by solving the scaled equations we can achieve approximately
$\mathcal{O}_{5}$ numerical precision, which is in comparison about two
orders of magnitude more accurate than the numerical solution of the
nonscaled equations of motion using the same time step.  Note that the 
$\tau$ scaling from Eq.~(\ref{scaling}) also affects the relaxation equations 
for the shear-stress tensor. Therefore the scaled quantities
$\tilde{\pi}^{\mu \nu }=\tau \pi ^{\mu \nu }$ result in
$d\tilde{\pi}^{\mu \nu }-\tilde{\pi}^{\mu \nu }/\tau =\tau d\pi^{\mu\nu}$.

For a better understanding of what will follow, we introduce the
notation $u^{\mu }=\gamma \left( 1,\mathbf{\bar{v}}_{i}\right) $ 
for the contravariant flow velocity from Eq.~(\ref{contravariant_u}).
Similarly, the covariant component is denoted as $u_{\mu }=\gamma \left( 1,-\text{%
\textbf{\b{v}}}_{i}\right) $; thus $v^{2}\equiv \mathbf{\bar{v}}_{i}$\textbf{%
\b{v}}$_{i}=v_{x}^{2}+v_{y}^{2}+\tau ^{2}v_{\eta }^{2}$ and $\gamma =\sqrt{%
1-v^{2}}$.

In our case of interest $g^{\tau \tau }=1$, and the metric of space-time
is diagonal leading to $g^{\tau i}=0$; thus we can introduce a simplified
notation which mimics the ideal fluid relations, $E\equiv T^{\tau \tau
}-\pi ^{\tau \tau }=T_{\tau \tau }-\pi _{\tau \tau }$, $\bar{M}_{i}\equiv
T^{\tau i}-\pi ^{\tau i}$, and \b{M}$_{i}\equiv T_{\tau i}-\pi _{\tau
i}=g_{\alpha \tau }g_{\beta i}\left( T^{\alpha \beta }-\pi ^{\alpha \beta
}\right) $. Using this notation we obtain the local rest frame energy density 
from Eqs.\ (\ref{T^00}) and~(\ref{T^0i}),%
\begin{equation}
e\equiv T^{\tau \tau }-\pi ^{\tau \tau }-(T^{\tau i}-\pi ^{\tau i})\text{%
\textbf{\b{v}}}_{i}=E-\bar{M}_{i}\text{\textbf{\b{v}}}_{i}\ ,  \label{e_LRF}
\end{equation}%
while the expression for the velocity components from Eq.~(\ref{T^0i}) leads
to, 
\begin{equation}
\mathbf{\bar{v}}_{i}\equiv \frac{T^{\tau i}-\pi ^{\tau i}}{T^{\tau \tau
}-\pi ^{\tau \tau }+P}=\frac{\bar{M}_{i}}{E+P}\,.  \label{v_i}
\end{equation}

Now, similarly expressing the $\mathbf{\text{\b{v}}}_{i}$ components
we define the magnitude of the three-velocity as
\begin{equation}
v\equiv \sqrt{\mathbf{\bar{v}}_{i}\text{\textbf{\b{v}}}_{i}}=\frac{M}{E+P}\,,
\label{v_magnitude}
\end{equation}%
where $M\equiv \left(\bar{M}_{i}\text{\b{M}}_{i}\right)^{1/2} =
\sqrt{\bar{M}_{x}^{2}+\bar{M}_{y}^{2}+\tau^{2}\bar{M}_{\eta}^{2}}$. Using
the latter two equations together we obtain
\begin{equation}
\mathbf{\bar{v}}_{i}=v\frac{\bar{M}_{i}}{M}.  \label{v_components}
\end{equation}%
Therefore with the help of Eq.~(\ref{e_LRF}), Eq.~(\ref{v_magnitude}) can be
solved using a one-dimensional root search, whereas Eq.~(\ref{v_components})
yields the individual velocity components.

In general we can reduce the number of unknowns in the relaxation
equations (\ref{relax_shear}) by applying the orthogonality and
tracelessness conditions of the shear-stress tensor. For example, by
choosing $\pi^{xx}$, $\pi^{yy}$, $\pi^{xy}$, $\pi^{x\eta }$, and
$\pi^{y\eta }$ as independent components, the other four components of the
shear-stress tensor follow from the orthogonality $\pi^{\mu \nu} u_{\nu} =0$,
\begin{eqnarray}
\pi ^{\tau \tau } &=&\pi ^{\tau x}v_{x}+\pi ^{\tau y}v_{y}+\tau ^{2}\pi
^{\tau \eta }v_{\eta }\,,  \label{pi_tautau} \\
\pi ^{\tau x} &=&\pi ^{xx}v_{x}+\pi ^{xy}v_{y}+\tau ^{2}\pi ^{x\eta }v_{\eta
}\,, \\
\pi ^{\tau y} &=&\pi ^{xy}v_{x}+\pi ^{yy}v_{y}+\tau ^{2}\pi ^{y\eta }v_{\eta
}\,, \\
\pi ^{\tau \eta } &=&\pi ^{x\eta }v_{x}+\pi ^{y\eta }v_{y}+\tau ^{2}\pi
^{\eta \eta }v_{\eta }\,,
\end{eqnarray}%
whereas the last unknown component is available from the tracelessness
condition $\pi^{\mu \nu} g_{\mu \nu} =0$:
\begin{align}
\pi ^{\eta \eta }& \equiv \tau ^{-2}\left( \pi ^{\tau \tau }-\pi ^{xx}-\pi
^{yy}\right)  \notag \\
& =\tau ^{-2}\left[ \pi ^{xx}\left( v_{x}^{2}-1\right) +\pi ^{yy}\left(
v_{y}^{2}-1\right) +2\pi ^{xy}v_{x}v_{y}\right.  \notag \\
& \left. +2\tau ^{2}\left( \pi ^{x\eta }v_{x}v_{\eta }+\pi ^{y\eta
}v_{y}v_{\eta }\right) \right] /\left( 1-\tau ^{2}v_{\eta }^{2}\right) \ .
\label{pi_etaeta}
\end{align}

Note that solving the above algebraic equations to obtain the
remaining five components, instead of explicitly propagating all ten
components of the shear-stress tensor, we introduce a small numerical
error compared to the latter method. This is because the velocities
entering into Eqs.~(\ref{pi_tautau})--(\ref{pi_etaeta}) are given from
the previous (half) time step, so we obtain different values with
different methods. However, this difference usually becomes smaller as
the number of time steps increases; hence this small numerical error
is acceptable especially if the runtime is also reduced considerably.

For sake of completeness we write out all terms from the
shear-stress relaxation equations explicitly. The relaxation equations
for the chosen five independent components of the shear-stress tensor
$\pi^{xx}$ , $\pi^{yy}$, $\pi^{xy} $, $\pi^{x\eta }$, and $\pi^{y\eta}$ are
\begin{align}
\tau _{\pi }d\pi ^{xx}& =2\eta _{s}\sigma ^{xx}-\pi ^{xx}-I^{xx}\ , \\
\tau _{\pi }d\pi ^{yy}& =2\eta _{s}\sigma ^{yy}-\pi ^{yy}-I^{yy}\ , \\
\tau _{\pi }d\pi ^{xy}& =2\eta _{s}\sigma ^{xy}-\pi ^{xy}-I^{xy}\ , \\
\tau _{\pi }d\pi ^{x\eta }& =2\eta _{s}\sigma ^{x\eta }-\pi ^{x\eta }  \notag
\\
& -\tau _{\pi }\frac{\gamma }{\tau }\left( \pi ^{x\eta }+v_{\eta }\pi ^{\tau
x}\right) -I^{x\eta }\ , \\
\tau _{\pi }d\pi ^{y\eta }& =2\eta _{s}\sigma ^{y\eta }-\pi ^{y\eta }  \notag
\\
& -\tau _{\pi }\frac{\gamma }{\tau }\left( \pi ^{y\eta }+v_{\eta }\pi ^{\tau
y}\right) -I^{y\eta }\ .\qquad
\end{align}

Here according to Eq.~(\ref{relax_shear}) we denoted
\begin{equation}
I^{\mu \nu }=I_{1}^{\mu \nu }+\delta _{\pi \pi }I_{2}^{\mu \nu }-\tau _{\pi
}I_{3}^{\mu \nu }+\tau _{\pi \pi }I_{4}^{\mu \nu }-\varphi _{7}I_{5}^{\mu
\nu }\ , 
\label{all_I_terms}
\end{equation}%
where 
\begin{align}
I_{1}^{\mu \nu }& =\left( \pi ^{\lambda \mu }u^{\nu }+\pi ^{\lambda \nu
}u^{\mu }\right) Du_{\lambda }\ , \\
I_{2}^{\mu \nu }& =\theta \pi ^{\mu \nu }\ , \\
I_{3}^{\mu \nu }& =\pi ^{\mu \lambda }\omega _{\left. {}\right. \lambda
}^{\nu }+\pi ^{\nu \lambda }\omega _{\left. {}\right. \lambda }^{\mu }\ , \\
I_{4}^{\mu \nu }& =\frac{1}{2}g_{\lambda \kappa }\left( \pi ^{\mu \kappa
}\sigma ^{\nu \lambda }+\pi ^{\nu \kappa }\sigma ^{\mu \lambda }\right) -%
\frac{1}{3}\Delta ^{\mu \nu }\pi _{\beta }^{\alpha }\sigma _{\alpha }^{\beta
}\ , \\
I_{5}^{\mu \nu }& =g_{\lambda \kappa }\pi ^{\mu \kappa }\pi ^{\nu \lambda }-%
\frac{1}{3}\Delta ^{\mu \nu }\pi _{\beta }^{\alpha }\pi _{\alpha }^{\beta }\
.
\end{align}%
The $I_{1}$ terms are%
\begin{align}
I_{1}^{xx}& =2\gamma v_{x}\left( \pi ^{\tau x}Du_{\tau }+\pi ^{xx}Du_{x}+\pi
^{yx}Du_{y}+\pi ^{\eta x}Du_{\eta }\right) \ , \\
I_{1}^{yy}& =2\gamma v_{y}\left( \pi ^{\tau y}Du_{\tau }+\pi ^{xy}Du_{x}+\pi
^{yy}Du_{y}+\pi ^{\eta y}Du_{\eta }\right) \ , \\
I_{1}^{xy}& =\gamma \left[ \left( \pi ^{\tau x}v_{y}+\pi ^{\tau
y}v_{x}\right) Du_{\tau }+\left( \pi ^{xx}v_{y}+\pi ^{xy}v_{x}\right)
Du_{x}\right.  \notag \\
& \left. +\left( \pi ^{yx}v_{y}+\pi ^{yy}v_{x}\right) Du_{y}+\left( \pi
^{\eta x}v_{y}+\pi ^{\eta y}v_{x}\right) Du_{\eta }\right] \ ,
\end{align}
\begin{align}
I_{1}^{x\eta }& =\gamma \left[ \left( \pi ^{\tau x}v_{\eta }+\pi ^{\tau \eta
}v_{x}\right) Du_{\tau }+\left( \pi ^{xx}v_{\eta }+\pi ^{x\eta }v_{x}\right)
Du_{x}\right.  \notag \\
& \left. +\left( \pi ^{yx}v_{\eta }+\pi ^{y\eta }v_{x}\right) Du_{y}+\left(
\pi ^{\eta x}v_{\eta }+\pi ^{\eta \eta }v_{x}\right) Du_{\eta }\right] \ , \\
I_{1}^{y\eta }& =\gamma \left[ \left( \pi ^{\tau y}v_{\eta }+\pi ^{\tau \eta
}v_{y}\right) Du_{\tau }+\left( \pi ^{xy}v_{\eta }+\pi ^{x\eta }v_{y}\right)
Du_{x}\right.  \notag \\
& \left. +\left( \pi ^{yy}v_{\eta }+\pi ^{y\eta }v_{y}\right) Du_{y}+\left(
\pi ^{\eta y}v_{\eta }+\pi ^{\eta \eta }v_{y}\right) Du_{\eta }\right] \ ,
\end{align}%
where according to Eqs.~(\ref{D_derivative}) and~\ref{d_derivative})
the proper time derivatives are given by 
$Du_{\mu } = du_{\mu}-\Gamma _{\mu \alpha }^{\beta }u^{\alpha }u_{\beta }$ and hence 
\begin{align}
Du_{\tau }& \equiv Du^{\tau } 
=\gamma \left[ \partial _{\tau }\gamma +v_{x}\partial _{x}\gamma
+v_{y}\partial _{y}\gamma +v_{\eta }\partial _{\eta }\gamma \right] 
 \notag \\ & +\tau
\gamma ^{2}v_{\eta }^{2}\ ,\  \\
Du_{x}& \equiv -Du^{x}=-\gamma \left[ \partial _{\tau }\left( \gamma
v_{x}\right) +v_{x}\partial _{x}\left( \gamma v_{x}\right) \right.  \notag \\
& \left. +v_{y}\partial _{y}\left( \gamma v_{x}\right) +v_{\eta }\partial
_{\eta }\left( \gamma v_{x}\right) \right] \ , \\
Du_{y}& \equiv -Du^{y}=-\gamma \left[ \partial _{\tau }\left( \gamma
v_{y}\right) +v_{x}\partial _{x}\left( \gamma v_{y}\right) \right.  \notag \\
& \left. +v_{y}\partial _{y}\left( \gamma v_{y}\right) +v_{\eta }\partial
_{\eta }\left( \gamma v_{y}\right) \right] \ , \\
Du_{\eta }& \equiv -\tau ^{2}Du^{\eta }=-\gamma \tau ^{2}\left[ \partial
_{\tau }\left( \gamma v_{\eta }\right) +v_{x}\partial _{x}\left( \gamma
v_{\eta }\right) \right.  \notag \\
& \left. +v_{y}\partial _{y}\left( \gamma v_{\eta }\right) +v_{\eta
}\partial _{\eta }\left( \gamma v_{\eta }\right) \right] -2\tau \gamma
^{2}v_{\eta }\ .
\end{align}%
Note that $Du_{\tau }\equiv du_{\tau }+\tau \gamma ^{2}v_{\eta
}^{2}=du^{\tau }+\tau \gamma ^{2}v_{\eta }^{2}$, $Du_{x}\equiv
du_{x}=-du^{x} $, $Du_{y}\equiv du_{y}=-du^{y}$, and $Du_{\eta }\equiv
du_{\eta }\neq du^{\eta }$, since $Du^{\eta }\equiv du^{\eta }+2\tau
^{-1}\gamma ^{2}v_{\eta }=-\tau ^{2}du_{\eta }$.

The $I_3$ terms are
\begin{align}
I_{3}^{xx}& =2\left( \pi ^{x\tau }\omega _{\left. {}\right. \tau }^{x}+\pi
^{xy}\omega _{\left. {}\right. y}^{x}+\pi ^{x\eta }\omega _{\left. {}\right.
\eta }^{x}\right) , \\
I_{3}^{yy}& =2\left( \pi ^{y\tau }\omega _{\left. {}\right. \tau }^{y}+\pi
^{yx}\omega _{\left. {}\right. x}^{y}+\pi ^{y\eta }\omega _{\left. {}\right.
\eta }^{y}\right) , \\
I_{3}^{xy}& =\pi ^{x\tau }\omega _{\left. {}\right. \tau }^{y}+\pi ^{y\tau
}\omega _{\left. {}\right. \tau }^{x}+\pi ^{xx}\omega _{\left. {}\right.
x}^{y}  \notag \\
& +\pi ^{yy}\omega _{\left. {}\right. y}^{x}+\pi ^{x\eta }\omega _{\left.
{}\right. \eta }^{y}+\pi ^{y\eta }\omega _{\left. {}\right. \eta }^{x}\ , \\
I_{3}^{x\eta }& =\pi ^{x\tau }\omega _{\left. {}\right. \tau }^{\eta }+\pi
^{\eta \tau }\omega _{\left. {}\right. \tau }^{x}+\pi ^{xx}\omega _{\left.
{}\right. x}^{\eta }  \notag \\
& +\pi ^{xy}\omega _{\left. {}\right. y}^{\eta }+\pi ^{\eta y}\omega
_{\left. {}\right. y}^{x}+\pi ^{\eta \eta }\omega _{\left. {}\right. \eta
}^{x}\ , \\
I_{3}^{y\eta }& =\pi ^{y\tau }\omega _{\left. {}\right. \tau }^{\eta }+\pi
^{\eta \tau }\omega _{\left. {}\right. \tau }^{y}+\pi ^{yx}\omega _{\left.
{}\right. x}^{\eta }  \notag \\
& +\pi ^{\eta x}\omega _{\left. {}\right. x}^{y}+\pi ^{yy}\omega _{\left.
{}\right. y}^{\eta }+\pi ^{\eta \eta }\omega _{\left. {}\right. \eta }^{y}\ ,
\end{align}%
where the vorticities are defined most generally as
\begin{align}
\omega _{\hspace*{0.1cm}\nu }^{\mu }& \equiv \frac{1}{2}\Delta ^{\mu \alpha
}\Delta _{\hspace*{0.1cm}\nu }^{\beta }\left( u_{\alpha ;\beta }-u_{\beta
;\alpha }\right)  \notag \\
& =\frac{1}{2}\left[ g^{\mu \alpha }\left( \partial _{\nu }u_{\alpha
}-u_{\nu }du_{\alpha }\right) -g_{\nu }^{\beta }\left( \partial ^{\mu
}u_{\beta }-u^{\mu }du_{\beta }\right) \right] \,  \notag \\
& +\frac{1}{2}\left( g^{\mu \alpha }u_{\nu }-g_{\hspace*{0.1cm}\nu }^{\alpha
}u^{\mu }\right) u^{\beta }\Gamma _{\alpha \beta }^{\lambda }u_{\lambda }.
\label{vorticity}
\end{align}
Here we used the fact that the Christoffel symbols of the second kind are
symmetric, $\Gamma_{\alpha \beta }^{\mu }=\Gamma _{\beta \alpha }^{\mu }$, with
respect to the interchange of the two lower indices.

The different components of the vorticity are given as
\begin{align}
\omega _{\left. {}\right. x}^{\tau }& \equiv \omega _{\left. {}\right. \tau
}^{x}=\frac{1}{2}\left[ \partial _{\tau }\left( \gamma v_{x}\right)
+\partial _{x}\gamma \right]   \notag \\
& +\frac{1}{2}\left[ \gamma v_{x}d\gamma -\gamma d\left( \gamma v_{x}\right) %
\right] +\frac{1}{2}\tau \gamma ^{3}v_{\eta }^{2}v_{x}\ , \\
\omega _{\left. {}\right. y}^{\tau }& \equiv \omega _{\left. {}\right. \tau
}^{y}=\frac{1}{2}\left[ \partial _{\tau }\left( \gamma v_{y}\right)
+\partial _{y}\gamma \right]   \notag \\
& +\frac{1}{2}\left[ \gamma v_{y}d\gamma -\gamma d\left( \gamma v_{y}\right) %
\right] +\frac{1}{2}\tau \gamma ^{3}v_{\eta }^{2}v_{y}\ , \\
\omega _{\left. {}\right. \eta }^{\tau }& \equiv \tau ^{2}\omega _{\left.
{}\right. \tau }^{\eta }=\frac{1}{2}\left[ \partial _{\tau }\left( \tau
^{2}\gamma v_{\eta }\right) +\partial _{\eta }\gamma \right]   \notag \\
& +\frac{1}{2}\left[ \tau ^{2}\gamma v_{\eta }d\gamma -\gamma d\left( \tau
^{2}\gamma v_{\eta }\right) \right] \ +\frac{1}{2}\tau ^{3}\gamma
^{3}v_{\eta }^{3}\ ,
\end{align}%
and%
\begin{align}
\omega _{\left. {}\right. y}^{x}& \equiv -\omega _{\left. {}\right. x}^{y}=%
\frac{1}{2}\left[ \partial _{y}\left( \gamma v_{x}\right) -\partial
_{x}\left( \gamma v_{y}\right) \right]   \notag \\
& +\frac{1}{2}\left[ \gamma v_{y}d\left( \gamma v_{x}\right) -\gamma
v_{x}d\left( \gamma v_{y}\right) \right] \ , \\
\omega _{\left. {}\right. \eta }^{x}& \equiv -\tau ^{2}\omega _{\left.
{}\right. x}^{\eta }=\frac{1}{2}\left[ \partial _{\eta }\left( \gamma
v_{x}\right) -\partial _{x}\left( \tau ^{2}\gamma v_{\eta }\right) \right]  
\notag \\
& +\frac{1}{2}\left[ \tau ^{2}\gamma v_{\eta }d\left( \gamma v_{x}\right)
-\gamma v_{x}d\left( \tau ^{2}\gamma v_{\eta }\right) \right] \ , \\
\omega _{\left. {}\right. \eta }^{y}& \equiv -\tau ^{2}\omega _{\left.
{}\right. y}^{\eta }=\frac{1}{2}\left[ \partial _{\eta }\left( \gamma
v_{y}\right) -\partial _{y}\left( \tau ^{2}\gamma v_{\eta }\right) \right]  
\notag \\
& +\frac{1}{2}\left[ \tau ^{2}\gamma v_{\eta }d\left( \gamma v_{y}\right)
-\gamma v_{y}d\left( \tau ^{2}\gamma v_{\eta }\right) \right] \ .
\end{align}

Note that the general expression of the vorticity given in Eq.~(10) in
Ref.~\cite{Molnar:2009tx} is missing the contribution of the
Christoffel symbols compared to Eq.~(\ref{vorticity}) in this
work. Therefore, the values for $\omega_{\left. {}\right. x}^{\tau}$,
$\omega_{\left. {}\right. y}^{\tau }$, and
$\omega_{\left. {}\right. \eta }^{\tau }$ given in Eqs.~(C.22), (C.23)
and (C.24) in Ref.~\cite{Molnar:2009tx} are also incorrect
compared to these formulas.

The next term we need is given by%
\begin{align}
I_{4}^{xx}& =\left( \pi ^{x\tau }\sigma ^{x\tau }-\pi ^{xx}\sigma ^{xx}-\pi
^{xy}\sigma ^{xy}-\tau ^{2}\pi ^{x\eta }\sigma ^{x\eta }\right)  \notag \\
& +\frac{1}{3}\left( 1+\gamma ^{2}v_{x}^{2}\right) \pi _{\beta }^{\alpha
}\sigma _{\alpha }^{\beta }\ , \\
I_{4}^{yy}& =\pi ^{y\tau }\sigma ^{y\tau }-\pi ^{yx}\sigma ^{yx}-\pi
^{yy}\sigma ^{yy}-\tau ^{2}\pi ^{y\eta }\sigma ^{y\eta }  \notag \\
& +\frac{1}{3}\left( 1+\gamma ^{2}v_{y}^{2}\right) \pi _{\beta }^{\alpha
}\sigma _{\alpha }^{\beta }\ , \\
I_{4}^{xy}& =\frac{1}{2}\left( \pi ^{x\tau }\sigma ^{y\tau }+\pi ^{y\tau
}\sigma ^{x\tau }\right) -\frac{1}{2}\left( \pi ^{xx}\sigma ^{yx}+\pi
^{yx}\sigma ^{xx}\right)  \notag \\
& -\frac{1}{2}\left( \pi ^{xy}\sigma ^{yy}+\pi ^{yy}\sigma ^{xy}\right) -%
\frac{\tau ^{2}}{2}\left( \pi ^{x\eta }\sigma ^{y\eta }+\pi ^{y\eta }\sigma
^{x\eta }\right)  \notag \\
& +\frac{1}{3}\left( \gamma ^{2}v_{x}v_{y}\right) \pi _{\beta }^{\alpha
}\sigma _{\alpha }^{\beta }\ , \\
I_{4}^{x\eta }& =\frac{1}{2}\left( \pi ^{x\tau }\sigma ^{\eta \tau }+\pi
^{\eta \tau }\sigma ^{x\tau }\right) -\frac{1}{2}\left( \pi ^{xx}\sigma
^{\eta x}+\pi ^{\eta x}\sigma ^{xx}\right)  \notag \\
& -\frac{1}{2}\left( \pi ^{xy}\sigma ^{\eta y}+\pi ^{\eta y}\sigma
^{xy}\right) -\frac{\tau ^{2}}{2}\left( \pi ^{x\eta }\sigma ^{\eta \eta
}+\pi ^{\eta \eta }\sigma ^{x\eta }\right)  \notag \\
& +\frac{1}{3}\left( \gamma ^{2}v_{x}v_{\eta }\right) \pi _{\beta }^{\alpha
}\sigma _{\alpha }^{\beta }\ , \\
I_{4}^{y\eta }& =\frac{1}{2}\left( \pi ^{y\tau }\sigma ^{\eta \tau }+\pi
^{\eta \tau }\sigma ^{y\tau }\right) -\frac{1}{2}\left( \pi ^{yx}\sigma
^{\eta x}+\pi ^{\eta x}\sigma ^{yx}\right)  \notag \\
& -\frac{1}{2}\left( \pi ^{yy}\sigma ^{\eta y}+\pi ^{\eta y}\sigma
^{yy}\right) -\frac{\tau ^{2}}{2}\left( \pi ^{y\eta }\sigma ^{\eta \eta
}+\pi ^{\eta \eta }\sigma ^{y\eta }\right)  \notag \\
& +\frac{1}{3}\left( \gamma ^{2}v_{y}v_{\eta }\right) \pi _{\beta }^{\alpha
}\sigma _{\alpha }^{\beta }\ .
\end{align}

The shear tensor is most generally defined as 
\begin{align}
\sigma^{\mu\nu} & \equiv \nabla^{\langle\mu}u^{\nu\rangle}=\frac{1}{2}\Delta^{\mu
\alpha }\Delta ^{\nu \beta }(u_{\alpha ;\beta }+u_{\beta ;\alpha })-\frac{%
\theta }{3}\Delta ^{\mu \nu }  \notag \\
& =\frac{1}{2}\left[ g^{\mu \alpha }\left( \partial ^{\nu }u_{\alpha
}-u^{\nu }du_{\alpha }\right) +g^{\nu \beta }\left( \partial ^{\mu }u_{\beta
}-u^{\mu }du_{\beta }\right) \right]  \notag \\
& -\Delta ^{\mu \alpha }\Delta ^{\nu \beta }\Gamma _{\alpha \beta }^{\lambda
}u_{\lambda }-\frac{\theta }{3}\Delta ^{\mu \nu },
\end{align}%
whereas the expansion scalar is
\begin{align}
\theta & \equiv \nabla _{\mu }u^{\mu }=\partial _{\mu }u^{\mu }+\Gamma
_{\lambda \mu }^{\lambda }u^{\mu }  \notag \\
& =\frac{\gamma }{\tau }+\partial _{\tau }\gamma +\partial _{x}\left( \gamma
v_{x}\right) +\partial _{y}\left( \gamma v_{y}\right) +\partial _{\eta
}\left( \gamma v_{\eta }\right) \ .
\end{align}%
The various shear tensor components that we need to use are
\begin{align}
\sigma ^{\tau \tau }& =-\tau \gamma ^{3}v_{\eta }^{2}+\left[ \left( \partial
_{\tau }\gamma -\gamma d\gamma \right) \right] +\left( \gamma ^{2}-1\right) 
\frac{\theta }{3}\ , \\
\sigma ^{\tau x}& =-\frac{1}{2}\left( \tau \gamma ^{3}v_{\eta
}^{2}v_{x}\right) +\frac{1}{2}\left[ \partial _{\tau }\left( \gamma
v_{x}\right) -\partial _{x}\gamma \right]  \notag \\
& -\frac{1}{2}\left[ \gamma v_{x}d\gamma +\gamma d\left( \gamma v_{x}\right) %
\right] +\gamma ^{2}v_{x}\frac{\theta }{3}\ , \\
\sigma ^{\tau y}& =-\frac{1}{2}\left( \tau \gamma ^{3}v_{\eta
}^{2}v_{y}\right) +\frac{1}{2}\left[ \partial _{\tau }\left( \gamma
v_{y}\right) -\partial _{y}\gamma \right]  \notag \\
& -\frac{1}{2}\left[ \gamma v_{y}d\gamma +\gamma d\left( \gamma v_{y}\right) %
\right] +\gamma ^{2}v_{y}\frac{\theta }{3}\ , \\
\sigma ^{\tau \eta }& =-\frac{\gamma ^{3}v_{\eta }}{2\tau }\left( 2+\tau
^{2}v_{\eta }^{2}\right) +\frac{1}{2}\left[ \partial _{\tau }\left( \gamma
v_{\eta }\right) -\frac{1}{\tau^2}\partial _{\eta }\gamma \right]  \notag \\
& -\frac{1}{2}\left[ \gamma v_{\eta }d\gamma +\gamma d\left( \gamma v_{\eta
}\right) \right] +\gamma ^{2}v_{\eta }\frac{\theta }{3}\ , \\
\sigma ^{\eta \eta }& =-\frac{\gamma }{\tau ^{3}}\left( 1+2\tau ^{2}\gamma
^{2}v_{\eta }^{2}\right) -\frac{1}{\tau ^{2}}\partial _{\eta }\left( \gamma
v_{\eta }\right)  \notag \\
& -\left( \gamma v_{\eta }\right) d\left( \gamma v_{\eta }\right) +\left( 
\frac{1}{\tau ^{2}}+\gamma ^{2}v_{\eta }^{2}\right) \frac{\theta }{3}\ ,
\end{align}%
and 
\begin{align}
\sigma ^{xx}& =-\left[ \partial _{x}\left( \gamma v_{x}\right) +\gamma
v_{x}d\left( \gamma v_{x}\right) \right] +\left( 1+\gamma
^{2}v_{x}^{2}\right) \frac{\theta }{3}\ , \\
\sigma ^{yy}& =-\left[ \partial _{y}\left( \gamma v_{y}\right) +\gamma
v_{y}d\left( \gamma v_{y}\right) \right] +\left( 1+\gamma
^{2}v_{y}^{2}\right) \frac{\theta }{3}\ , \\
\sigma ^{xy}& =-\frac{1}{2}\left[ \partial _{x}\left( \gamma v_{y}\right)
+\partial _{y}\left( \gamma v_{x}\right) \right]  \notag \\
& -\frac{1}{2}\left[ \gamma v_{y}d\left( \gamma v_{x}\right) +\gamma
v_{x}d\left( \gamma v_{y}\right) \right] +\gamma ^{2}v_{x}v_{y}\frac{\theta 
}{3}\ , \\
\sigma ^{x\eta }& =-\frac{\gamma ^{3}v_{x}v_{\eta }}{\tau }-\frac{1}{2}\left[
\partial _{x}\left( \gamma v_{\eta }\right) +\frac{1}{\tau ^{2}}\partial
_{\eta }\left( \gamma v_{x}\right) \right]  \notag \\
& -\frac{1}{2}\left[ \gamma v_{\eta }d\left( \gamma v_{x}\right) +\gamma
v_{x}d\left( \gamma v_{\eta }\right) \right] +\gamma ^{2}v_{x}v_{\eta }\frac{%
\theta }{3}\ , \\
\sigma ^{y\eta }& =-\frac{\gamma ^{3}v_{y}v_{\eta }}{\tau }-\frac{1}{2}\left[
\partial _{y}\left( \gamma v_{\eta }\right) +\frac{1}{\tau ^{2}}\partial
_{\eta }\left( \gamma v_{y}\right) \right]  \notag \\
& -\frac{1}{2}\left[ \gamma v_{\eta }d\left( \gamma v_{y}\right) +\gamma
v_{y}d\left( \gamma v_{\eta }\right) \right] +\gamma ^{2}v_{y}v_{\eta }\frac{%
\theta }{3}\ .
\end{align}%
The last contribution from Eq.~(\ref{all_I_terms}) are
\begin{align}
I_{5}^{xx}& =\left( \pi ^{x\tau }\right) ^{2}-\left( \pi ^{xx}\right)
^{2}-\left( \pi ^{xy}\right) ^{2}-\left( \tau \pi ^{x\eta }\right) ^{2} 
\notag \\
& +\frac{1}{3}\left( 1+\gamma ^{2}v_{x}^{2}\right) \pi _{\beta }^{\alpha
}\pi _{\alpha }^{\beta }\ , \\
I_{5}^{yy}& =\left( \pi ^{y\tau }\right) ^{2}-\left( \pi ^{yx}\right)
^{2}-\left( \pi ^{yy}\right) ^{2}-\left( \tau \pi ^{y\eta }\right) ^{2} 
\notag \\
& +\frac{1}{3}\left( 1+\gamma ^{2}v_{y}^{2}\right) \pi _{\beta }^{\alpha
}\pi _{\alpha }^{\beta }\ , \\
I_{5}^{xy}& =\pi ^{x\tau }\pi ^{y\tau }-\pi ^{xx}\pi ^{yx}-\pi ^{xy}\pi
^{yy}-\tau ^{2}\pi ^{x\eta }\pi ^{y\eta }  \notag \\
& +\frac{1}{3}\left( \gamma ^{2}v_{x}v_{y}\right) \pi _{\beta }^{\alpha }\pi
_{\alpha }^{\beta }\ , \\
I_{5}^{x\eta }& =\pi ^{x\tau }\pi ^{\eta \tau }-\pi ^{xx}\pi ^{\eta x}-\pi
^{xy}\pi ^{\eta y}-\tau ^{2}\pi ^{x\eta }\pi ^{\eta \eta }  \notag \\
& +\frac{1}{3}\left( \gamma ^{2}v_{x}v_{\eta }\right) \pi _{\beta }^{\alpha
}\pi _{\alpha }^{\beta }\ , \\
I_{5}^{y\eta }& =\pi ^{y\tau }\pi ^{\eta \tau }-\pi ^{yx}\pi ^{\eta x}-\pi
^{yy}\pi ^{\eta y}-\tau ^{2}\pi ^{y\eta }\pi ^{\eta \eta }  \notag \\
& +\frac{1}{3}\left( \gamma ^{2}v_{y}v_{\eta }\right) \pi _{\beta }^{\alpha
}\pi _{\alpha }^{\beta }\ .
\end{align}

Furthermore, to evaluate the Cooper-Frye formula, Eq.~(\ref{Cooper-Frye}),
as well as the argument of the equilibrium distribution function,
Eq.~(\ref{eq_distribution}), we express the four-momenta of particles as,
\begin{equation}
p^{\mu }=\left( m_{T}\cosh \left( y_{p}-\eta \right) ,p_{x},p_{y},\frac{m_{T}%
}{\tau }\sinh \left( y_{p}-\eta \right) \right) ,
\end{equation}%
where $m$ is the rest mass of the particle,
$m_{T}=\sqrt{m^{2}+p_{x}^{2}+p_{y}^{2}}$ denotes the transverse mass,
while $y_{p}$ is the rapidity of the particle.  Therefore, the
nonequilibrium corrections to the spectra from Eq.~(\ref{f_final})
are given with an argument of
\begin{align}
& \pi ^{\alpha \beta }p_{\alpha }p_{\beta }  \notag \\
& =m_{T}^{2}\left[ \cosh ^{2}\left( y_{p}-\eta \right) \pi ^{\tau \tau
}+\tau ^{2}\sinh ^{2}\left( y_{p}-\eta \right) \pi ^{\eta \eta }\right] 
\notag \\
& +\left( p_{x}^{2}\pi ^{xx}+2p_{x}p_{y}\pi ^{xy}+p_{y}^{2}\pi ^{yy}\right) 
\notag \\
& -2m_{T}\cosh \left( y_{p}-\eta \right) \left( p_{x}\pi ^{\tau x}+p_{y}\pi
^{\tau y}\right)  \notag \\
& +2\tau m_{T}\sinh \left( y_{p}-\eta \right) \left( p_{x}\pi ^{x\eta
}+p_{y}\pi ^{y\eta }\right)  \notag \\
& -2\tau m_{T}^{2}\sinh \left( y_{p}-\eta \right) \cosh \left( y_{p}-\eta
\right) \pi ^{\tau \eta },
\end{align}
while using Eq.~(\ref{d_sigma_tau_eta}) we obtain
\begin{align}
p^{\mu }d^{3}\Sigma _{\mu }& = \tau\left[ m_{T}\cosh \left( y_{p}-\eta \right)dxdyd\eta
-p_{x}d\tau dyd\eta  \right. \notag \\
& \left. - p_{y}d\tau dxd\eta  -\frac{m_{T}}{\tau }\sinh \left( y_{p}-\eta \right) d\tau dxdy\right]  .
\end{align}%

\section{Numerical methods}

\label{shasta}

The conservation laws as well as the relaxation equations are solved
using the well known SHASTA ''SHarp and Smooth Transport Algorithm''
originally developed by Boris and Book~\cite{Borisetal} and later
refined by Zalesak~\cite{Zalesak} and others \cite{FCT_book}. This
numerical algorithm solves equations of the conservation type with
source terms:
\begin{equation}
\partial_t U + \partial_i (v_i U) = S(t,\mathbf{x}) \, ,  \label{eq:generic}
\end{equation}
where $U = U(t,\mathbf{x})$ is for example $T^{00}$ or $T^{0i}$, while
$v_i $ is the $i$th component of three-velocity, and $S(t,\mathbf{x})$
is a source term; for more details see
Refs.~\cite{Muronga:2006zw,Molnar:2008fv,Molnar:2009tx}.

Because for smooth solutions (like in our case) the multidimensional
antidiffusion limiter suffers from instabilities around the boundary
caused by small ripples propagating into the interior
\cite{Kuzmin_2004}, we further stabilized SHASTA by letting the
antidiffusion coefficient $A_{ad}$, which controls the amount of
numerical diffusion, be proportional to
\begin{equation}
A_{ad} = \frac{A^S_{ad}}{\left(k/e\right)^2 + 1},
\end{equation}
where $A^S_{ad}=0.125$ is the default value for the antidiffusion
coefficient \cite{Borisetal}, $e$ is the energy density in the local
rest frame, and $k=6\times 10^{-5}$ GeV/fm$^3$ is a numerical
parameter. In this way we increase the amount of numerical diffusion
in the low-density region and $A_{ad}$ goes smoothly to zero near the
boundaries of the grid. In our cases of interest this neither affects
the solution nor produces more entropy inside the decoupling surface.

It is also important to mention that in the (3+1)-dimensional case
both the conservation and the relaxation equations are solved using
SHASTA, employing the above-mentioned modification for the
antidiffusion coefficient. Earlier, for the (2+1)-dimensional
boost-invariant case, we used a simple centered second-order
difference algorithm to solve the relaxation equations
\cite{Niemi:2012ry}. However doing so in the (3+1)-dimensional case
does not always lead to stable solutions.

To further stabilize the numerical calculations (and also for ideal
fluids) we used a smaller value for the antidiffusion coefficient in
the transverse directions, $A^{x,y}_{ad}=0.105$, but kept
$A^\eta_{ad}=0.125$ in the $\eta$ direction.  Decreasing the
antidiffusion coefficient produces smoother solutions inside the
decoupling hypersurface but also increases the numerical diffusion,
which in turn may decrease the numerical accuracy.  The reason we used
a different coefficient in the longitudinal direction is to increase
the accuracy; see the next section for more details.

The numerical calculations are done on a discretized spatial grid
(including four boundary points in each direction as required by the
algorithm) of $N_{x}\times N_{y}\times N_{\eta }$ cells with
$N_{x}=N_{y}=180$ while $N_{\eta }=2\times 120$ with $\Delta x=\Delta
y=\Delta \eta =0.15$ fm cell sizes. The time step is given from
$\Delta \tau =\lambda \Delta x$, which for $\lambda =0.4$ leads to
$\Delta \tau =0.06 $ fm/$c$. Furthermore, the system is symmetric around
the $x$ and $y$ directions, with exponentially interpolated
boundary conditions for the conserved quantities (\emph{e.g.},\ for
Glauber-type initial conditions) and linearly interpolated boundary
conditions for the shear-stress tensor (because the shear-stress tensor may
change sign).

Finally, the freeze-out hypersurface is constructed at time intervals
$\Delta \tau_{CF} \equiv 5 \Delta \tau = 0.3$ fm/$c$. The space is
sampled uniformly in both the transverse and longitudinal
directions, at $\Delta x_{CF} \equiv 2 \Delta x = 0.3$ fm distances.

The freeze-out hypersurface is calculated using the
\textsc{cornelius++} subroutine presented in
Ref.~\cite{Huovinen:2012is} and its source code can be obtained from
the Open Standard Codes and Routines (OSCAR) website \cite{cornelius}.

\section{Remarks on the numerical accuracy}
\label{shasta_accuracy}

SHASTA solves the fluid dynamical equations up to some finite
numerical accuracy. In most cases this means that in Cartesian
coordinates the particle number and energy are conserved up to
$\mathcal{O}_{5}$ accuracy. However in $\left( \tau ,x,y,\eta \right)$
coordinates the expressions for the conserved quantities as well as
the equations of motion change with additional source terms resulting
from the nonvanishing Christoffel symbols.

As an example let us evaluate a conserved quantity at a given time or
proper time; hence by comparing this initial value with one at a later
time we can follow the accuracy of the fluid-dynamical solver during
this time interval.

The total conserved charge $N_{tot}$ across any given hypersurface is%
\begin{equation}
N_{tot}\equiv \int N^{\mu }d^{3}\Sigma _{\mu }=\int N^{0}d^{3}\Sigma
_{0}+\int N^{i}d^{3}\Sigma _{i}.  \label{N_tot}
\end{equation}%
Here the hypersurface element $d^{3}\Sigma_{\mu }$ can be specified in any
coordinate system according to the following general formula:
\begin{equation}
d^{3}\Sigma _{\mu }=-\epsilon _{\mu \nu \lambda \kappa }\frac{\partial
\Sigma ^{\nu }}{\partial u}\frac{\partial \Sigma ^{\lambda }}{\partial v}%
\frac{\partial \Sigma ^{\kappa }}{\partial w}dudvdw\,,
\end{equation}%
where $\epsilon _{\mu \nu \lambda \kappa }$ is the Levi-Civita symbol.

For example in Cartesian coordinates the hypersurface normal vector is $%
\Sigma _{\mu }^{\left( t,z\right) }\left( t,x,y,z\right)$, where $t=t\left(
x,y,z\right) $; hence 
\begin{eqnarray}
d^{3}\Sigma _{\mu }^{\left( t,z\right) } &\equiv &\left(
dxdydz,-dtdydz,-dtdxdz,-dtdxdy\right)  \notag \\
&=&\tau \left( \frac{\partial \tau }{\tau \partial \eta }\sinh \eta +\cosh
\eta ,-\frac{\partial \tau }{\partial x},-\frac{\partial \tau }{\partial y}%
,\right.  \notag \\
&&\left. -\frac{\partial \tau }{\tau \partial \eta }\cosh \eta -\sinh \eta
\right) dxdyd\eta ,
\end{eqnarray}%
while in $\left( \tau ,x,y,\eta \right) $ coordinates for $\Sigma _{\mu
}^{\left( \tau ,\eta \right) }\left( \tau ,x,y,\eta \right) $ and $\tau
=\tau \left( x,y,\eta \right) $ we obtain 
\begin{equation}
d^{3}\Sigma _{\mu }^{\left( \tau ,\eta \right) }=\tau \left( dxdyd\eta
,-d\tau dyd\eta ,-d\tau dxd\eta ,-d\tau dxdy\right) .
\label{d_sigma_tau_eta}
\end{equation}

If we are interested in the conserved current across constant time or
proper time hypersurfaces then $d^{3}\Sigma _{i}^{\left( t,z\right)
}=d^{3}\Sigma _{i}^{\left( \tau ,\eta \right) }=0$; hence in Cartesian
coordinates we get, 
\begin{equation}
N_{tot}\left( t\right) \equiv \int N^{\mu }d^{3}\Sigma _{\mu }^{\left(
t,z\right) }=\gamma n_{0}\int dxdydz,
\end{equation}%
where $N^{\mu }\equiv n_{0}u^{\mu }=\gamma n_{0}\left(
1,v_{x},v_{y},v_{z}\right) $ is the conserved charge current.
Similarly, Eq.~(\ref{N_tot}) leads to the total conserved charge at any
proper-time hypersurface in hyperbolic coordinates, 
\begin{equation}
N_{tot}\left( \tau \right) \equiv \int N^{\mu }d^{3}\Sigma _{\mu }^{\left(
\tau ,\eta \right) }=\gamma n_{0}\int \tau dxdyd\eta .
\end{equation}%

To calculate how the total energy-momentum changes between two closed
hypersurfaces, first we define the energy-momentum current across a
hypersuface as
\begin{equation}
E_{tot}^{\mu }\equiv \int T^{\mu \nu }d^{3}\Sigma _{\nu }=\int T^{\mu
0}d^{3}\Sigma _{0}+\int T^{\mu i}d^{3}\Sigma _{i}.  \label{E_total_current}
\end{equation}%
In Cartesian coordinates $E_{tot}^{\mu }=\left(
E_{tot}^{0},E_{tot}^{i}\right) $, such that $E_{tot}^{0}$ denotes the energy
current while $E_{tot}^{i}$ denotes the momentum current trough the hypersurface.
Therefore the total energy current across a constant-$t$ hypersurface is 
\begin{equation}
E_{tot}^{0}\left( t\right) \equiv \int T^{0\nu }d^{3}\Sigma _{\nu }^{\left(
t,z\right) }=\int T^{00}dxdydz.  \label{E_tot_t_z}
\end{equation}

The energy-momentum current across a constant-$\tau$
hypersurface in $\left( \tau ,x,y,\eta \right) $ coordinates can also be
calculated from Eq.~(\ref{E_total_current}) together with the general
transformation rules $E_{tot}^{\mu }=\left( \partial x^{\mu }/\partial \hat{x%
}^{\alpha }\right) \hat{E}_{tot}^{\alpha }$, where the position vectors are $%
x^{\mu }\equiv \left( t,x,y,z\right) =\left( \tau \cosh \eta ,x,y,\tau \sinh
\eta \right) $ and $\hat{x}^{\mu }\equiv \left( \tau ,x,y,\eta \right) $.
Thus the total energy across a constant-$\tau$ hypersurface is given by 
\begin{align}
E_{tot}^{0}\left( \tau \right) & \equiv \int \cosh \eta\; T^{\tau \nu
}d^{3}\Sigma _{\nu }^{\left( \tau ,\eta \right) }+\int \tau \sinh \eta\;
T^{\eta \nu }d^{3}\Sigma _{\nu }^{\left( \tau ,\eta \right) }  \notag \\
& =\int \left( \cosh \eta\; T^{\tau \tau }+\tau \sinh \eta\; T^{\eta \tau
}\right) \tau dxdyd\eta .  \label{E_tot_tau_eta}
\end{align}

Using the latter formulas we can check energy conservation from the
initial time to the end using
\begin{equation}
\Delta E_{tot}^{0} \left( t \right)= E_{tot}^{0}\left( t_{end}\right)
-E_{tot}^{0}\left( t_{ini}\right).
\end{equation}%
It turns out that by solving the fluid dynamical equations in Cartesian
coordinates we can achieve $\Delta E_{tot}^{0}(t) \approx \mathcal{O}_{6}$
numerical accuracy, while in hyperbolic coordinates 
$\Delta E_{tot}^{0}(\tau) \approx \mathcal{O}_{1}$. This behavior is due to
two different reasons.

First, the numerical algorithm is accurate only to finite
precision, meaning that $T^{00}$ or $T^{\tau \tau }$ is calculated
correctly only up to the first six digits. However, due to the
hyperbolic functions in Eq.~(\ref{E_tot_tau_eta}) the total energy of
the system is given by a differently weighted sum over all cells
(compared to Cartesian coordinates).  These hyperbolic weights
increase very rapidly as a function of $\eta$; hence even though the
numerical error of the solver is acceptably small for SHASTA, the
weighted sum over all cells in hyperbolic coordinates shows otherwise.

We have checked that for RHIC energies $\Delta E_{tot}^{0}(\tau) < 2\%$
while at LHC energies this number can be as much as $20\%$. This
is because $f\left( \eta \right) $ is much narrower at RHIC
than at the LHC.  Similar results were also obtained in
Ref.~\cite{Karpenko:2013wva} using a different computational
fluid-dynamical algorithm.

We also verified energy conservation inside the constant-temperature
freeze-out hypersurface, and we found that in that case the energy is
conserved at $1\%$ accuracy, at both RHIC and the LHC. This was
expected since inside the $T=100$ MeV freeze-out hypersurface the
space-time rapidity of matter is $\eta < 10$.


\end{document}